
\input phyzzx
%
%
%
\countdef\dfnumber=80
\countdef\thmnumber=81
\count80=1
\count81=1
\def\no{\noindent {\bf Note.\ }}
\def\thm{\noindent {\bf Theorem \number\thmnumber. }
\advance\thmnumber by 1}
\def\prf{\noindent {\bf Proof. }}
\def\df{\noindent {\bf Definition \number\dfnumber. } \advance\dfnumber by 1}
\def\o{\omega}
\def\s{\sigma}
\def\O{{\cal O}}
\def\ie{{\it i.e.}\ }
\def\L{{\cal L}}
\def\q{{\kappa}}
\def\e{e_{-\psi}}
\def\T{{\rm T}_\q \O}
\def\S{{\rm S}^1}
\def\C{C^\infty}
\def\V{{\cal V}_\q}
\def\a{\alpha}
\def\b{\beta}
\def\c{\gamma}
\def\d{\delta}

\def\Ad{{\rm Ad}^*}
\def\A{{\rm Ad}}
\def\ad{{\rm ad}^*}
\def\mod{{\rm \ mod\ }}
\def\Ker{{\rm Ker}}
\def\Ann{{ \rm Ann}}
\def\inv{$\Ad$-invariant\ }
\def\sphere{{\bf S}^1}

\def\deriv{{d \over d \epsilon} \Big|_{\epsilon=0}}
%
%

\Ref\abraham{R. Abraham, J.E. Marsden.: Foundations of Mechanics.
Second Edition. Benjamin/ Cummings Pub. 1980.}

\Ref\adler{M. Adler.: On the Trace Functional for Formal Pseudo-Differential
Operators and the Symplectic Structure of the Korteweg-Devries Type Equations.
Inv. Math. 50, 219-248 (1979).}

\Ref\babelon{O. Babelon, C.M. Viallet.: Integrable Models,
Yang-Baxter Equation and Quantum Groups. Part 1. Preprint SISSA-54/89/EP May
1989.}

\Ref\bakas{I. Bakas, D.A. Depireux.:
A Fractional KdV Hierarchy, Mod. Phys. Lett. A6, 1561 (1991).
(Erratum A6, 2351 (1991)).}

\Ref\burr{N. Burroughs, M. De Groot, T. Hollowood, J.L. Miramontes.:
Generalised Drinfel'd-Sokolov Hierarchies II: The Hamiltonian Structures.
Preprint PUPT-1263, IASSNS-HEP-91/42.}

\Ref\DJ{E. Date, M. Jimbo, M. Kashiwara, T. Miwa.: Transformation
Groups for Soliton Equations-Euclidean Lie Algebras and Reduction
of the KP Hierarchy. Publ. RIMS. Kyoto Univ. 18, 1077-1110 (1982).}

\Ref\Tim{M. De Groot, T.J. Hollowood, J.L. Miramontes.: Generalized
Drinfel'd-Sokolov Hierarchies. IAS and Princeton preprint IASSNS-HEP-91/19,
PUPT-1251 March 1991. To appear in CMP.}

\Ref\DS{V.G. Drinfel'd, V.V. Sokolov.: Lie Algebras and Equations of the
 Korteweg-de Vries
Type. Jour.Sov.Math. {\bf 30} (1985) 1975;
Equations of Korteweg-De Vries Type and
Simple Lie Algebras. Soviet.Math.Dokl. {\bf23} (1981) 457.}

\Ref\FT{L.D. Faddeev, L.A. Takhtajan.: Hamiltonian Methods
in the Theory of Solitons. Springer-Verlag, 1986.}

\Ref\kaccam{V.G. Kac.: Infinite Dimensional Lie Algebras, $2^{nd}$ edition.
Cambridge University Press 1985.}

\Ref\KP{V.G. Kac, D.H. Peterson.: 112 Constructions of the Basic
Representations of the Loop Group of $E_8$. In Symposium
on Anomalies, Geometry and Topology. Eds: W. A. Bardeen \& A. R. White.
Singapore, World Scientific 1985.}

\Ref\Jim{M. Jimbo, T. Miwa.: Solitons and Infinite Dimensional Lie Algebras.
Publ. RIMS. Kyoto Univ. {\bf 19}, 943-1001 (1983).}

\Ref\kirillov{A.A. Kirillov.: Elements of the Theory of Representations.
Springer-Verlag 1976.}

\Ref\LP{V.F. Lazutkin, T.F. Pankratova.: Normal Forms and Versal
Deformations for Hill's Equation. Func. Anal. \& Appl. {\bf 9},
306-311 (1975).}

\Ref\MS{L.J. Mason, G.A.J. Sparling.: Nonlinear Schro\"odinger and
Korteweg-de-Vries are
reductions of self-dual Yang-Mills. Phys.Lett. {\bf A137} (1989) 29.}

\Ref\MW{J. Marsden, A. Weinstein.: Reduction of
Symplectic Manifolds with Symmetry. Rep. Math. Phys. {\bf 5}, 121-130, 1974.}

\Ref\segal{G. Segal.: Unitary Representations of some infinite dimensional
groups. CMP {\bf 80}, 301-342, (1981).}

\Ref\Segal{G. Segal.: The Geometry of the KdV equation. Int. Jour.
Mod. Phy. A. Vol. 6 No 16, 2859-2869, (1991).}

\Ref\SW{G. Segal, G. Wilson.: Loop Groups and Equations of KdV Type.
IHES Pub. Math, No {\bf 61}, 5-65 (1985).}

\Ref\shansky{M. Semenov-Tian-Shansky.: Dressing Transformations and
Poisson Group Actions. Publ. RIMS. Kyoto Univ. {\bf 21} (1985) 1237.}

\Ref\weinstein{A. Weinstein.: The Local structure of
Poisson Manifolds. Jour. Diff. Geom. {\bf 18}, 523-557 (1983).}

\date={October 9$^{th}$, 1991}
\Pubnum{IASSNS-HEP-91/67}
\titlepage
\title{COADJOINT ORBITS OF THE GENERALISED
$Sl(2),\ Sl(3)$ KdV HIERARCHIES}

\author{Nigel J. Burroughs}
\address{Institute For Advanced Study,\break
Olden Lane, Princeton, N.J. 08540}

\abstract{In this paper we develop two coadjoint orbit constructions
for the phase spaces of the generalised $Sl(2)$ and $Sl(3)$
KdV hierachies.
This involves the construction of two group actions
in terms of Yang Baxter operators,
and an Hamiltonian reduction of the coadjoint orbits.
The Poisson brackets are reproduced by the Kirillov construction.
{}From this construction we obtain a `natural' gauge fixing
proceedure for the generalised hierarchies.}

\endpage

\baselineskip=14pt
\tenpoint

\chapter{Introduction}

This paper analyses the initial steps
in a coadjoint orbit construction for the generalised KdV hierarchies,
[\burr,\Tim].
The analysis proceeds principally through the medium of illustration,
using the KdV hierarchies constructed on the Kac Moody algebras $\hat
{sl}(2)$ and $\hat {sl}(3)$. In this paper we prove that for these
theories there exist two
orbit constructions for the phase space, reproducing
via the Kirillov construction, [\kirillov], the
two Poisson brackets constructed in [\burr].

The {\it Coadjoint Orbit Method} (also known as the
Adler-Kostant-Symes formalism, or AKS formalism)
is a construction that uses Lie algebras to define integrable models,
[\babelon,\FT].
The essential ingredients are a Lie algebra $g$, and an endomorphism
$R : g \rightarrow g$ satisfying the {\it modified Yang Baxter
Equation}, mYBE.
The mYBE implies that the bracket $\left[X,Y \right]_R=
\left[RX,Y \right]+ \left[X, RY \right]$ satisfies the Jacobi
identity, and thus defines a second Lie bracket on the Lie algebra
$g$. The
Kirillov construction for  Poisson brackets on the dual Lie
algebra $g^*$, [\kirillov], defines two Poisson brackets $\{\, ,\, \},
\{\, ,\, \}_R$ on $g^*$ induced from $[\, ,\, ]$ and $[\, ,\, ]_R$
respectively.
The fact that the Poisson bracket $\{\, ,\, \}_R$ is constructed from a
group action allows the symplectic leaves of $\{\, , \,\}_R$ to be
constructed, each symplectic leaf, $\O$, furnishing a phase space for a
dynamical system with the symplectic structure $\{\, ,\, \}_R |_\O$.
Further, a set of commuting Hamiltonians can be
constructed, these being the \inv functions.
Thus there exist a set of Hamiltonians $\{H_i \}$ satisfying
$\{H_i, H_j \}_R=0$ which generate, under Poisson brackets,
a set of commuting
time flows.
The fact that these theories are integrable follows from the existence
of sufficient commuting Hamiltonians.
The aspect of this method that concerns us in this paper is the
construction of the symplectic leaves of $\{\, , \, \}_R$. These are
the {\it coadjoint} orbits of the group $G_R$ on $g^*$, where $G_R$ denotes
the exponential of the Lie algebra $g$ with commutator $[\, ,\, ]_R$.
If there exists an inner product on $g$, then the dual algebra can be
identified with the Lie algebra. Thus the coadjoint orbits
are identified with subspaces of $g$.
In the case of a current algebra with the Schwinger central extension,
the theory is of Lax type, with a Lax operator
of the form $\L=\partial_x +\q$, where $\q \in \O \subset g$. The
equations of motion take the form of the zero curvature condition,
$\left[\partial_t+M_i, \partial_x+\q \right]=0$,
where $M_i$ is related to the
functional derivative of the Hamiltonian $H_i$ generating the time
coordinate $t_i$, $M_i=R \left( d_\q H_i \right)$.

We observe that the KdV hierarchies of [\Tim,\DS] have many
features reminiscient of this coadjoint orbit construction. In
particular, the theories are constructed by exploiting the structure
of a Lie algebra, \ie the current algebra on a loop algebra $\hat g$;
the Poisson structures are expressed in terms of Lie brackets involving
R-operators, [\burr]; and the equations of motion take the form of the
zero curvature equations. In addition, there exist coadjoint orbit
formulations for the Toda Chain, [\babelon], and Non-linear
Schr\"odinger equation, [\FT], both special cases of the generalised
KdV-hierarchies, [\Tim]. These are special cases because they have no
gauge group, and only possess a single Poisson structure.
The KdV hierarchies are in general  bi-Hamiltonian, [\burr],
and it is for this reason that
a coadjoint orbit construction may be inappropriate
in describing  these theories;
there appears to be no known method to
extend the AKS proceedure to create bi-Hamiltonian systems.
In this paper, we attempt to initiate a bi-Hamiltonian construction
from AKS systems. The idea is to construct two Lie algebra commutators
on the current algebra $C^\infty(\sphere, \hat g)$, denoted
$[\, ,\, ]_R $ and $[\, ,\, ]_\s$, such that the Poisson brackets of
the KdV hierarchies, [\burr], are reproduced
by the Kirillov bracket construction: $\{\, ,\, \}_\s=\{\, ,\, \}_1,
\{\, ,\, \}_R=\{\, ,\, \}_2$.
We denote the `exponential' of these Lie algebras by $G_R, G_\s$ respectively.
If we perform the AKS process, we would obtain two integrable
systems with phase spaces $\O_R$ and $\O_\s$, coadjoint
orbits of the groups $G_R$ and $G_\s$ respectively, and a set of
commuting Hamiltonians that are identical for both theories.
The idea is that the
gauge symmetry of the KdV hierarchy
is the additional ingredient that
equates the two theories dynamically.
We perform an
Hamiltonian reduction on the two orbits  such that the
reduced phases spaces become identified. Further, the symplectic
structures are inequivalent, leading to a bi-Hamiltonian structure.

This is in fact a simplification of the process.
Our final conclusion is that the
reduced phase space of $\O_\s$ is identical to the phase space of a
generalised KdV hierarchy, for an appropriate choice of orbit and gauge group.
However, the reduced orbits of the group $G_R$
foliate this phase space, \ie under the Poisson bracket
$\{\, ,\, \}_R$ the
phase space of the generalised KdV hierarchy is not symplectic, and
breaks it into symplectic leaves, [\weinstein], leaves that can be
reproduced as Hamiltonian reductions of $G_R$-orbits.
The flows of the hierarchy are such that
this foliation is preserved, and thus there is no inconsistency.
This foliation induces a partition of the potentials of the hierarchy
into `types', part
of this partition reproducing the distinction between mKdV type, and
`true' KdV type potentials.

The coadjoint orbit structure proposed here should be contrasted with
the construction for the $sl(2)$-KdV hierarchy as a coadjoint
orbit of a central extension of
$Diff^+(\sphere)$, [\LP,\segal, \Segal]. This differs from the orbit
structures considered in this paper, because we do not consider
reparametrisations of $\sphere$. This structure is specific to the
case of $sl(2)$, not generalising to more general hierarchies.
We further comment that a coadjoint orbit construction of the
traditional $A_n$-KdV hierarchies exists within the framework of
Pseudo-Differential Operators, [\adler]. Orbits of a group of formal
pseudo-differential symbols are constructed, the second
Poisson bracket $\{\, ,\, \}_2$ of the hierarchy being reproduced as the orbit
symplectic structure.
Since the generalised hierarchies do not appear to possess a
description in terms of pseudo-differential operators, this orbit
structure cannot be generalised to these cases.

This paper is organised as follows. In section 2 we review the theory
of the momentum map and the theory of Hamiltonian reduction.
We specialise this discussion to
the case of the  Hamiltonian
reduction of a coadjoint orbit. In section 3 we review the content of
[\burr] and the definition of the two Poisson brackets of the
generalised KdV hierarchies. We propose, in section 4, a
construction
of the KdV hierarchies as the Hamiltonian reduction of
a coadjoint orbit of the  group  $G_\s$. We further propose
that an Hamiltonian reduction of the coadjoint orbits of the $G_R$-action
are capable of describing the dynamics, providing an
explanation for the existence of the
two Poisson structures.
In section 5 we construct the momentum maps for the two gauge groups,
$H_R, H_\s$,
the symmetry groups to be used in the Hamiltonian reduction of the
orbits $\O_R, \O_\s$ of the traditional $A_n$-hierarchies.
In section 6, we discuss the $G_R$-action, and the
reduction by the gauge group $\C(\sphere, N_-)$.
In the following two
sections we analyse the traditional $Sl(2), Sl(3)$ KdV hierarchies as
coadjoint orbit systems. In section  9 we extend the momentum
map analysis to include the case of the fractional
KdV hierarchies, [\bakas], and generalised hierarchies of [\Tim].
In section 10, we discuss the reduction
of the $G_R$-orbits for  these theories, the symmetry group of $\O_R$
being a subgroup of $\C(\sphere, N_-)$ in general, and not
$\C(\sphere, N_-)$ itself.
We use these
results in our two further examples involving the
following non-traditional choices
for $\Lambda$,
$$
\Lambda_\alpha=\pmatrix{0&0&1 \cr 0&0&0 \cr z&0&0 \cr},
\ \Lambda_{{\rm co}}^2=\pmatrix{0&0&1 \cr z&0&0 \cr 0&z&0}.
$$

Throughout this paper we shall not treat the difficulties involved
in infinite dimensional phase spaces, assuming that the finite results
generalise without difficulty.

\chapter{HAMILTONIAN REDUCTION: GENERAL THEORY}

In this section we summarise the salient features of momentum mappings and
Hamiltonian reduction. This exposition follows that of [\abraham].

Consider a symplectic manifold $(P, \o)$ with group action $\Phi : G \times P
\rightarrow P$, such that $\Phi_g$ is a symplectomorphism
for all $g \in G$. The momentum map is defined as follows:

\df
{\it The momentum mapping is a map $J : P \rightarrow g^*$
such that the Hamiltonian functions $\hat J_X$ defined by
$$
\hat J_X (x)= \langle J(x), X \rangle, \ \forall X \in g, x \in P,
\eqn\hamfunc
$$
generate the action $\Phi$ under Poisson brackets,} \ie
$$
d \hat J_X = {\bf i}_X \o,\ {\rm where\ } \left(X, {\bf i}_Y \o \right)
=\left( X \wedge Y, \o \right), \ \forall X, Y \in TP.
\eqn\momgen
$$

Note that here we are using the embedding $g \rightarrow TP$ induced from
the infinitesimal action of $G$ on $P$,
\ie we use the same symbol $X$ to denote $X \in g$ and the vector
field
$Xf(x)={d \over d \epsilon} \big|_{\epsilon=0} f(\Phi(e^{\epsilon X}, x))$.
Equation \momgen\ corresponds to using the isomorphism $T_x P \cong
T^*_x P$ induced by $\omega$ to map the vector $X$ to the
corresponding 1-form.

Given a symplectic action $\Phi : G \times P \rightarrow P$ there may not
exist a momentum mapping. The obstruction lies in solving the equation
\momgen\ globally. If a solution $\hat J_X$ exists for all $X \in g$ then a
momentum mapping exists, and the group action is generated
by Poisson brackets $ \delta_X \phi(x)={d \over d \epsilon} \big|_{\epsilon=0}
\phi \left(\Phi\left({\rm
exp} \epsilon X, x\right) \right) = \{\hat J_X, \phi \}(x)$, using \momgen.
Under Poisson brackets the Lie algebra $g$ may be centrally extended.
This depends on the equivariance relation of the momentum map.

\df
{\it The equivariance relation of the momentum map
$J$ is the commutation relation}:
$$
J\left(\Phi_g(x)\right)= \Psi_{g}\cdot J(x),\ \forall x \in P, g \in
G,
$$
{\it where $\Psi$ is a group action of $G$ on $g^*$ defined by
a cohomology class $\s \in H^1(G,g^*)$}
$$
\Psi : (g,l) \rightarrow \Ad(g) \cdot l +\s(g),\,
\forall l \in g^*,g \in G.
$$
{\it The momentum map of an action $G \times P \rightarrow P$ is
classified by the cohomology class  $\s \in H^1(G,g^*)$.}

\no
Given a symplectic action $\Phi$, the momentum map can be calculated by
solving for the Hamiltonian functions $\hat J_X$ from equation \momgen.
If a momentum map exists
globally, we can calculate the equivariance relation
from the formula $\s(g)=J(\Phi_g (x))- \Ad(g) \cdot J(x)$ for any
point $x \in P$.

The importance of the cocycle $\s$ is that the Poisson bracket algebra
is centrally extended:

\thm
{\it Given a momentum map $J : P \rightarrow g^*$ classified by the
cohomology class $\s \in H^1(G, g^*)$, the Lie algebra $g$ is centrally
extended under Poisson brackets}
$$
\{\hat J_X, \hat J_Y \}= \hat J_{[X,Y]}+ \Sigma (X,Y),\
{\rm where\ } \Sigma (X,Y) = \deriv \langle \s ({\rm exp}(\epsilon X), Y
\rangle.
$$

We now consider the Hamiltonian reduction by a symmetry
$\Phi : G \times P \rightarrow P$. Assume that there exists a momentum map
$J : P \rightarrow g^*$ with an equivariance relation involving a
cohomology class $\s \in H^1(G, g^*)$. Then the Hamiltonian reduction
involves two processes. First we restrict to a submanifold $J^{-1}(l_0)
\subset P$, where $l_0 \in g^*$ is a regular value of $J$.
These submanifolds are the {\it level
sets} of the momentum map $J$, and correspond to fixing the values of
the constants of motion $\hat J_X, \forall X \in g$. Then we take
equivalence classes under the {\it little group} of the point
$l_0$; $G(l_0)=\{g \in G \, |\, \Psi (g) \cdot l_0 =l_0 \}$. The reduced phase
space is thus $P_{{\rm red}}= J^{-1}(l_0) / G(l_0)$. General theory implies
that there is a symplectic structure on $P_{{\rm red}}$ induced by
this process, the corresponding 2-form being denoted $\omega_{{l_0}}$.
The 2-form $\omega_{{l_0}}$
is related to the original 2-form $\omega$ by the
relation, [\MW],
$\pi^*_{{l_0}} \omega_{{l_0}}=i^*_{{l_0}} \omega$, where
$\pi_{{l_0}}$ is the projection $\pi_{{l_0}} :
J^{-1}(l_0) \rightarrow P_{{\rm red}}$, and $i_{{l_0}}$ is the
inclusion $i_{{l_0}} : J^{-1}(l_0) \rightarrow P$.
This relation also holds in the infinite dimensional case
under certain assumptions.
The construction of the reduced manifold $P_{{\rm red}}$
can be considered as a
method for constructing a symplectic leaf of the  Poisson manifold
$P \big/ G$, the space of $\Phi$-orbits of $G$ in $P$. Since the action
$\Phi$ is symplectic, the space $ P \big/ G$ inherits a Poisson
structure, \ie if $f, l$ are functions on $P \big/ G$ and $\pi : P
\rightarrow P \big/ G$, then the Poisson structure satisfies
$\{ \pi^*f, \pi^* l \}= \pi^* \{f, l\}$ which implies that it is well
defined on $G$-invariant functions. The momentum defined by $l_0$
selects a symplectic leaf in $P \big/ G$. We observe that in this
framework, the fact that $P$ is symplectic is not necessary, \ie we
can reduce a Poisson manifold by a symmetry, selecting a symplectic
leaf as the phase space.

Comparing to a more familiar reduction process, we observe that
for a system with translation invariance in a direction $z$,
restricting to the level set corresponds to fixing the $p_z$ momentum,
while taking equivalence classes
under the little group corresponds to eliminating
the $z$ coordinate.

We observe that the image of the moment map $J(P) \subset g^*$ is
necessarily a symplectic submanifold of $g^*$, endowed with the Kirillov
bracket constructed from the extended group action $\Psi$. This implies
that the image is in fact an orbit of the $\Psi$ action of $G$ on $g^*$.

Our interest is in the Hamiltonian reduction of a coadjoint orbit of a Lie
group $G$.
Consider a
coadjoint orbit $\O= \Ad_R (G) \cdot \Lambda \subset g^*,\, \Lambda \in g^*$
of a Lie group
$G$ on
the dual Lie algebra $g^*$, with the group action defined in terms of
a classical Yang Baxter operator $R$, [\babelon].
Suppose we are reducing with respect to a Lie
group $H$, with Lie algebra $h$, action $\Phi$ and momentum map $J$. Then
equation \momgen\ can be rewritten as
$$
\ad_R (d_\q \hat J_Y) \cdot \q = \deriv \Phi({\rm exp} \epsilon Y, \q),
\forall Y \in h, {\rm \ and \ } \q \in \O.
\eqn\momorbit
$$
The fact that the orbit $\O$ is symplectic implies that this equation
is consistent locally, \ie the only problem in solving this
equation is  the exactness of the 1-form $d_\q \hat J_Y$ as before.

The tangent space of the orbit $\O$ at a point $\q \in \O$ is
generated by the Lie algebra $g$ through $\ad_R$ action, \ie $T_\q\O
={\rm span}_{X \in g}( \ad_R(X) \cdot \q )$.
Correspondingly, the
tangent space of the level set of $J$
is given by restricting $g$ to a subspace $S_\q \subset g$ given by
$$
S_\q=\{ X \in g\, \big| \,
\langle \ad_R(X) \cdot \q, d \hat J_Y \rangle=0\ \forall Y \in h\}.
\eqn\Sdefinition
$$
This is simply the requirement that the
Hamiltonians $\hat J_Y$ are constant on the level sets.
The tangent space of the level set is given by ${\rm span}_{X \in S_\q} (
\ad_R(X) \cdot \q )$.
If we define $\V$ as the set of directions
that are generated infinitesimally by $\Phi$ through the point $\q \in
\O$, \ie
$$
\V=\Bigl\{ r \in T_\q \O \, \Bigl|\,
\deriv \Phi \left({\rm exp} \epsilon Y, \q \right)
=\q+\epsilon r+ O(\epsilon^2), \ {\rm for\ some\ } Y \in H\Bigr\},
$$
then the subset $S_\q$ can be rewritten as
$$
S_\q=\{X \in g \, |\, \langle X, \V \rangle =0 \}
\eqn\leveltangents
$$
This follows from a rearrangement of \Sdefinition,
$$
\langle \ad_R(d \hat J_Y) \cdot \q, X \rangle
\equiv \langle  \deriv \Phi \left({\rm exp} \epsilon Y, \q \right),X
\rangle=0.
$$
This means that the set $S_\q$ is orthogonal to
the directions generated by  the
infinitesimal action of $H$.
Under the equivalence relation with the little group $G(l_0), l_0= J(\q)$,
the tangent
space at the equivalence class $\q / G(l_0)$ is represented by
the set
$$
\ad_R (X) \cdot \q \mod \V, \forall X \in S_\q.
\eqn\likegauge
$$

Observe that the restriction to the subspace $S_\q$, \Sdefinition, is very
similar to the gauge invariance constraint
on the functional derivatives in the theory of
the KdV hierarchy, [\burr, \DS]. Further, the equivalence in \likegauge\ is
very similar to the gauge equivalence also employed.

\chapter{REVIEW}

In this section we review the content of [\burr].
In [\burr],
two Poisson structures
are constructed for the generalised KdV hierarchies. This construction
was restricted to those hierachies
constructed on untwisted Kac Moody algebras.

The central object in the construction of the hierarchies
is a Kac-Moody algebra $\hat
g$, realized as the loop algebra $\hat g=g\otimes{\bf
C}[z,z^{-1}]\oplus{\bf C}d$, where $g$ is a finite Lie algebra with Lie
group $G$.
The
derivation $d$ is chosen to induce the {\it homogeneous\/} gradation, so
that $[d,a\otimes z^n]=n\,a\otimes z^n$ $\forall\,a\in g$.
We use the following notation :
$\{e_i\}_{i=1}^{{\rm rank}(g)}$
are the raising operators associated to the simple roots
of $g$ in a Cartan-Weyl basis of $g$; $\psi$ is the highest root, $\e$
the corresponding lowering operator;
$N_\pm, B_\pm$ are the Borel subgroups of $G$;
$k_i$ are the Kac Labels of $g$, and $[w]$ denotes the conjugacy class
of the Weyl reflection $w$ of the root space of $g$.
There are other gradations on $\hat g$ given by,
[\kaccam]

\df
{\it A  gradation of type ${\bf s}$, is defined
via the derivation $d_{\bf s}$ which satisfies
$$[d_{\bf s},e_i\otimes z^n]=(nN+s_i)e_i\otimes z^n,$$
where $N=\sum_{i=0}^{{\rm
rank}(g)}k_i s_i$ and ${\bf
s}=(s_0,s_1,\ldots,s_{{\rm rank}(g)})$ is a vector of rank$(g)+1$ non-negative
integers.}

Under a gradation of type $\bf
s$, $\hat g$ is a ${\bf Z}$-graded algebra:
$$\hat g=\bigoplus_{i\in{\bf Z}}\hat g_i({\bf s}).$$
The homogeneous gradation corresponds to ${\bf s}_{\rm hom}\equiv(1,0,\ldots,
0)$.
A special class of gradations are those constructed from
conjugacy classes of the Weyl group of $g$, [\kaccam].
Associated to a conjugacy
class $[w]$, there exists a gradation denoted ${\bf s}[w]$. The
conjugacy classes of the Weyl group also classify the Heisenberg
subalgebras, [\KP], the Heisenberg subalgebra corresponding to $[w]$
being denoted ${\cal H}[w]$.
The Coxeter
class, $[w_{{\rm co}}]$, defines the familiar {\it principal}
gradation, and corresponds to ${\bf s}=(1,1,...1)$.
There exists a partial ordering on the set of gradations
of $\hat g$, [\Tim], given by : ${\bf s}
\preceq {\bf s}^\prime$ if ${\bf s}_i \not= 0$ whenever ${\bf
s}^\prime_i \not=0$.
The construction of the hierarchies involves the use of two
gradations, one of which is induced from a conjugacy class $[w]$.
These are denoted ${\bf s}$ and ${\bf s}[w]$.
To reduce the complexity of the notation, we employ the following
notation to refer to them. We use subscripts to denote
$\bf s$-grade, and superscripts to denote ${\bf s}[w]$-grades,
\ie $\hat g_j\equiv\hat g_j({\bf s})$ and
$\hat g^j\equiv\hat g_j({\bf s}[w])$.
In this paper
our interest is principally in the KdV-type hierarchies,
constructed from the homogeneous gradation ${\bf s}={\bf s}_{{\rm hom}}$.

The KdV hierarchies are constructed from the data
$(\Lambda,{\bf s},[w])$, where ${\bf s} \preceq {\bf s}[w]$ and
$\Lambda$ is a (constant) element of $\hat g$ with ${\bf s}[w]$-grade $i>0$.
{}From this data,
one defines the Lax operator
$$
\L=\partial_x+q+\Lambda.
\eqn\Aa
$$
The {\it potential} $q$ is
defined to be an element of $C^\infty(\sphere,\hat g_{\geq 0} \cap
\hat g^{<i})$.
The potentials are taken to be periodic functions, this avoiding
technical complications [\DS].
The function $q(x)$ plays the r\^ole of the phase space coordinate in
this system. However, there exist symmetries
in the system corresponding to the gauge transformation
$\L\rightarrow S\L S^{-1}$,
with $S$ generated by $x$ dependent functions on the
subalgebra $\hat g_0 \cap \hat g^{<0}$.
The {\it phase
space\/} of the system ${\cal M}$ is the set of gauge equivalence
classes of operators of the form $\L=\partial_x+q+\Lambda$.
The space of functions ${\cal F}$ on ${\cal M}$ is the
set of gauge invariant
functionals of $q$ of the form
$$\varphi[q]=\int_{{\bf R}/{\bf Z}}
dx\,f\left(x,q(x),q^\prime(x),\ldots,q^{(n)}(x),\ldots\right).$$
It is straightforward to find a basis for ${\cal F}$, the gauge
invariant functionals. One simply performs a non-singular
gauge transformation to take $q$ to some canonical form $q^{\rm can}$.
The components of $q^{\rm can}$ and their derivatives then provide the desired
basis.

The outcome of applying the procedure of Drinfel'd and Sokolov, [\DS], to \Aa\
is
that there exists an infinite number of commuting flows on the gauge
equivalence classes of $\L$, [\Tim].
In [\burr] it is further proved that these flows are bi-Hamiltonian,
\ie there exist two coordinated
Poisson brackets on the phase space $\cal M$ such that
the flows are generated by commuting Hamiltonians.
These Hamiltonians are labeled by elements of the Heisenberg algebra ${\cal
H}[w]$, [\Tim], \ie $H_b$ denotes the Hamiltonian defined by the
constant element $b \in {\cal H}[w]$ with ${\bf s}[w]$-grade $>0$.
It is important that
the hierarchy is constructed with a gradation induced
from a conjugacy class $[w]$ because the existence of the Heisenberg
subalgebra ${\cal H}[w]$ is essential to the construction.
We extract the
following theorem from [\burr]

\thm
{\it There is a one parameter family of Hamiltonian structures on
the gauge equivalence classes of the generalized KdV hierarchy given
by\/}
$$
\{\varphi,\psi\}_\mu=\left(q+\Lambda,\left[d_q\varphi,d_q\psi\right]_{R_\mu}
\right)-\left(d_q\varphi,\left(d_q\psi\right)^\prime\right),
$$
{\it where $[\ ,\ ]_{R_\mu}$ is the Lie algebra commutator
constructed from $R_\mu=(P_{\geq0}-P_{<0})/2-\mu/z$.
Expanding in powers of $\mu$, $\{\ ,\ \}_\mu=\mu\{\ ,\ \}_1+\{\ ,\ \}_2$,
we obtain the
two coordinated Hamiltonian structures on ${\cal M}$\/}
$$\eqalign{
\{\varphi,\psi\}_1=&-\left(d_q\varphi,z^{-1}\left[
d_q\psi,\L\right]\right),\cr
\{\varphi,\psi\}_2=&\left(q+\Lambda,\left[d_q\varphi,d_q\psi\right]_{R}
\right)-\left(d_q\varphi,\left(d_q\psi\right)^\prime\right),\cr}
$$
{\it where $R=(P_{\geq0}-P_{<0})/2$.
Under time evolution in the coordinate $t_b$,
the following recursion relation holds\/}:

$$
{\partial\varphi\over\partial t_b}=
\{\varphi,H_{zb}\}_1=\{\varphi,H_b\}_2.
$$

\chapter{THE ORBITS}

In this section we outline a program to explain the previous results
in terms of a coadjoint orbit construction on the current algebra of
$\hat g$. This  construction will be further explained
in another publication.
The two Poisson brackets are interpreted as
Kirillov brackets by the construction of two Lie algebra commutators
on $\C(\sphere, \hat g)$ defined in terms of two
Yang Baxter operators, $R, R^\s$.
The bi-Hamiltonian
structure then follows if certain conditions are satisfied, in
particular the orbits being
dynamically equivalent under  Hamiltonian reduction.

Define the two Yang Baxter operators
$R=P_{\geq0}-P_{<0}$ and $R^\s=P_{>0}- P_{\leq0}$ on the Kac Moody
algebra $\hat g$. These define Lie algebra commutators on $\hat g$,
denoted by $[\, , \, ]_R$,
and $[\, , \, ]_{R^\s}$ respectively.
Our interest lies in the Lie bracket $[\, ,\, ]_R$ and
the shifted commutator
$[\, , \, ]_\s={1 \over z} [\, , \,]_{R^\s}$.
On the current algebra,
the commutator $[\, ,\, ]_R$ is centrally extended with the two form
$$
\omega(X,Y)= \int_{\sphere}dx \langle P_0(X(x)), Y(x)^\prime \rangle.
\eqn\schwingerext
$$
The Kirillov construction of Poisson brackets on the dual of a
Lie algebra,
[\kirillov], defines two Poisson brackets on $\C(\sphere, \hat g^*)$,
which are identical to those of the hierarchy, [\burr],
$\{\, ,\, \}_1= \{\, ,\, \}_\s, \{\, ,\, \}_2= \{\, ,\, \}_R$.
As in the traditional analysis of the coadjoint orbit method,
we induce the coadjoint
actions $\Ad_R$ and $\Ad_\s$ on the dual $C^\infty(\sphere, \hat g^*)$.
The coadjoint orbits of $\Ad_{R/\s}$
define the symplectic leaves of the Poisson
brackets $\{\, ,\, \}_R, \  \{\, ,\, \}_\s$ respectively. We know that
on each orbit we can construct an integrable system. The proposal in
this paper is that these orbits can be modified (by Hamiltonian reduction)
such that the
generalised KdV hierarchies are reproduced as dynamical systems, the
hierarchies constructed from $({\bf s}_{{\rm hom}}, {\bf s}[w])$ having a
description in terms of both $\Ad$ actions, thus reproducing the two
Poisson brackets of [\burr, \DS].

Before discussing the $\Ad_{R/ \s}$-orbits, we
analyse the relationship between the two Poisson structures on the
phase space. For a bi--Hamiltonian system, the dynamics are generated
through either Poisson bracket, with the
bi-Hamiltonian constraint
$$
\{\phi, H_b \}_2 = \{ \phi, H_{zb} \}_1
$$
for all Hamiltonians $H_b$, and functionals $\phi \in {\cal F}$.
It is only through this relation that the Poisson brackets are
related, thus the requirement that the theories are dynamically
equivalent does not in fact imply that the phase spaces are identical.
The condition for dynamical equivalence is
that the phase space of the theory is symplectic under
the Poisson bracket $\{\, ,\, \}_1$, and the Poisson bracket $\{\,
,\, \}_2$ produces a foliation of the phase space, this
foliation being preserved under all the flows. Since the flows of the
hierarchy are generated by the Poisson bracket $\{\, , \,\}_2$, the
requirement that the flows preserve the foliation is trivially satisfied.
Thus, in attempting to construct the generalised KdV hierarchies from the
Coadjoint Orbit Method, we should choose an orbit of $\Ad_\s$ as the
phase space and prove that this phase space is preserved by the
$\Ad_R$-action, thus proving the foliation requirement.
However, this is an over simplification of the construction, and will
in fact fail. This is because the phase spaces of the KdV hierarchies
are not  $\Ad_\s$ orbits. Thus we must modify our proposal. This
modification can be deduced from an analysis of the orbits, and
involves an Hamiltonian reduction of the orbits.

Identifying the dual with the original Lie algebra through the inner
product, the $\Ad_R, \Ad_\s$-orbits are identified as subspaces of
$\C(\sphere, \hat g)$ and
have the following structure

\noindent
{\it The $G_R$ action}
$$
\q \rightarrow
P_{\leq0} \left( g_{\geq0} \L g_{\geq0}^{-1} \right) +
P_{>0} \left( g_{<0}^{-1} \L g_{<0} \right)
\eqn\generalR
$$

This action preserves the decomposition $\q=\{\q_{>0},\q_{\leq0} \}$.

\noindent
{\it The $G_\s$ action}
$$
\q \rightarrow
P_{<0} \left( g_{\geq0} \L g_{\geq0}^{-1} \right) +
P_{\geq0} \left( g_{<0}^{-1} \L g_{<0} \right)
\eqn\generalsig
$$

This action preserves the decomposition $\q=\{\q_{\geq0},\q_{<0} \}$.

Here $g_{\geq0}$ denotes the formal
exponential of an element of the Lie
algebra $\C(\sphere,\hat g_{\geq0})$, and similarly for $g_{<0}$.
These actions can be made more precise through the use of a
representation, or the universal enveloping algebra.

Taking the hint from the KdV hierarchy, we simplify the group actions
by generating the orbits from a  point in
$\hat g_{\geq0}$. Both group actions
preserve the space $\hat g_{\geq0}$, which implies that the orbits also
lie in $\hat g_{\geq0}$.
These orbits take the form

\noindent
{\it The $G_R$-orbit}
$$
\q \rightarrow \left( g_{0} \L_0 g_{0}^{-1} \right)+
P_{>0} \left( g_{<0}^{-1} \L g_{<0} \right)
\eqn\orR
$$
Note that the second term only contributes if $ \L_{\geq 2} \not=0$,
\ie if the Lax operator possesses terms of homogeneous degree $>1$.
In this paper we restrict to the case $\L_2=0$, and hence the second
term reduces to $\L_1, \ \forall g_{<0}$.

\noindent
{\it The $G_\s$-orbit}
$$
\q \rightarrow P_{\geq0} \left( g_{<0}^{-1} \L g_{<0} \right)
\eqn\orsig
$$

Observe that both group actions treat $\hat g_{>0}$ identically.
However they differ in their treatment of the space $\hat g_0$.
Hence we require a mechanism that equates the dynamical degrees of
freedom in $\hat g_0$.
This process must preserve the symplectic nature of the orbits,
which suggests that we perform an Hamiltonian reduction with respect to a
symmetry group of each orbit.
Thus for each orbit, $\O_\s$, $\O_R$, we require a group of
symplectomorphisms, $H_\s$, $H_R$ respectively, with which to perform an
Hamiltonian reduction.
One class of symplectomorphisms that are easily identified are those
induced from an algebra homomorphism $\chi : \C(\sphere,\hat g)
\rightarrow \C(\sphere,\hat
g)$, $\chi$ being an algebra homomorphism with respect to both Lie
algebra structures $[\,,\,]_R$ and $[\,,\,]_\s$. Since these Lie brackets
depend on the homogeneous gradation, adjoint action by any element of
$G$ will
necessarily be an algebra homomorphism.
The symplectic submanifolds $\O_\s$ and $\O_R$ are in fact sufficiently similar
that an Hamiltonian reduction by subgroups $H_\s, H_R \subset G$
is sufficient,
where $H_\s, H_R$ act by Ad$^*$-action.
Note that one of the constraints on $H_\s, H_R$ is that under $\Ad$-action the
orbits are preserved, \ie we require
$\Ad (H_\s) \cdot \O_\s \subset \O_\s$,
$\Ad (H_R) \cdot \O_R \subset \O_R$.
Assume for the moment that these actions possess moment maps
$J_\s : \O_\s \rightarrow h^*_\s$, $J_R : \O_R \rightarrow h^*_R$,
where $h_\s, h_R$ are the Lie algebras of $H_\s,H_R$
respectively, and $h^*_{\s/R}$ denote the corresponding duals.
Then the Hamiltonian reduction is performed by restricting to
the inverse image $J_\s^{-1} (n_\s)$, and taking equivalence classes
under the little group $G_\s(n_\s)$, $n_\s \in h^*_\s$.
The condition that the reduced phase spaces are
equivalent dynamically now reduces to
the requirement that
$J^{-1}_\s (n_\s)/ G_\s(n_\s)$
is foliated by spaces of the form
$J_R^{-1} (n_R)/ G_R(n_R)$.
Further, we expect for appropriate choices of orbit $\O_\s$,
symmetry group $H_\s$ and momentum $n_\s$, that the reduced phase space is
identical to that of a generalised KdV hierarchy of [\Tim].

The preceeding discussion leads us to propose the following conjecture
for the integrable models constructed in [\Tim],

\noindent
{\bf Conjecture.}
{\it
The $( {\bf s}_{{\rm hom}}, {\bf s}[w] )$
KdV Hierarchies
have  phase spaces that are
Hamiltonian reductions of $G_\s$-orbits, by a symmetry group
$H_\s \subset \C(\sphere, N_-)$. The leaves of the
foliation induced by the second Poisson structure, $\{\, ,\, \}_R$, are
Hamiltonian reductions of $G_R$-orbits, with a symmetry group $H_R
\subset \C(\sphere, N_-)$.
This foliation
induces a partition of the potentials
that is finer than the
separation of the potentials into modified KdV,
partially modified KdV  and KdV type
potentials.}

We  observe that prior to the introduction of the gauge group
$H_\s$, the orbit $\O_\s$ is effectively finite since there is no
central extension in the $G_\s$-action.
It is only through the gauge group that the
$G_\s$-orbit acquires a complexity capable of describing theories such
as the generalised KdV hierarchies.

The first test of this conjecture is whether
it is able to reproduce the traditional $A_n$-KdV hierarchies, [\DS].
In these theories, the element $\Lambda$ has
${\bf s}[w_{{\rm co}}]$-grade equal to 1, and
the homogeneous decomposition $\Lambda=I+z \e$, where $I=
\sum_{i=1}^{{\rm rank}(g)} e_i$.
The orbits are constructed such that they pass through the point
$\q=\q_0+z \e \in C^\infty(\sphere,\hat g_{\geq0})$, \ie
initially we only fix
the component of homogeneous degree 1.
The orbits can now be written down explicitly:
$$\eqalign{
\O_R=& \{z \e+ g^{-1} \left( \partial_x +\q_0 \right) g \, |\,
\forall g \in \C(\sphere, G) \},\cr
\O_\s=& \{\q+ \left[\e,Y \right]
\, |\, \forall Y \in \C(\sphere,g \mod \Ker(\e))\},\cr}
$$
where we note that the $\O_\s$ orbit is parametrised by the space
$\C(\sphere, g \mod \Ker(\e))$, with $\Ker(X)=\{ Y \in g \, | \, [Y,X]=0 \}$.
To simplify the comparison of the
orbits, we write down the tangent spaces of the orbits at the
point $\q=\Lambda_{z \rightarrow z+\mu}$,
$$\eqalign{
T_{\Lambda+\mu \e} \O_R=& {\rm span}_{X \in \C(\sphere, g)}\left(
\mu \left[\e,X \right]+ \left[\partial_x+I,X \right] \right), \cr
T_{\Lambda+\mu \e} \O_\s=& {\rm span}_{X \in \C(\sphere, g)}\left(
\left[\e,X \right] \right).\cr}
\eqn\tangentspan
$$
All elements of the tangent space of $\O_\s$
are orthogonal to Ker$(\e)$.

Observe that the two tangent spaces are similar, the choice
$\q=\Lambda+\mu \e$ in \tangentspan\ eccentuating the similarity.
However
the tangent spaces $T_\q\O_R$ and $T_\q \O_\s$ of the two
orbits are not equivalent,
in particular the $G_R$-orbit has
directions that are not orthogonal to $\Ker(\e)$. Further,
there are additional degrees of freedom over and above those
in $g \mod \Ker(\e)$, corresponding to the terms $\left[ \partial_x+I,X
\right]$. Thus the $\O_\s$ orbit is more restricted than the $\O_R$
orbit as a model for a phase space.
Hamiltonian reduction
is able to reduce this discrepancy between the tangent spaces,
\ie we could enforce $\T_R$ to be
orthogonal to $\Ker(\e) \subset g$ through the introduction of a symmetry.
However, this is too strong a condition, since the introduction of a
symmetry also introduces an equivalence relation on directions in the
tangent space, \likegauge.
Thus we only impose a symmetry by a subgroup
of the Lie group generated by $\C(\sphere,\Ker(\e))$.
Since the tangent space $\T_\s$ can never have a direction in the
upper triangular subalgebra $n_+$ for any point in the orbit,
the gauge algebra $h_R$ must at least include $\C(\sphere, n_-)$ as a
subalgebra. From the known structure of the KdV hierarchy, this should
in fact be sufficient.

\chapter{THE GAUGE GROUP MOMENTUM MAPS AND HAMILTONIAN REDUCTION}

In this section, for the case when
$\Lambda$ has ${\bf s}[w_{{\rm co}}]$-grade 1,
we construct the two momentum maps $J_R,\, J_\s$ for the
$\Ad$-action of $H_R, H_\s \subset \C(\S,G)$ on the orbits $\O_\s$ and $\O_R$.
The moment map $J_R$ is a simple quotient, while the moment map $J_\s$
requires a rather complex calculation. After calculating the
equivariance relation of $J_\s$, we find that under Poisson
brackets the Lie algebra $h_\s$ is centrally extended.

Given a symmetry $\Phi : G \times P \rightarrow P$, the
momentum map is calculated by solving equation \momgen, or more
specifically \momorbit\ in the case of a coadjoint orbit.
This corresponds to infinitesimally generating the action $\Phi$
under Poisson brackets. For the gauge group actions,
$H_R, H_\s \in \C(\S,G)$, this equation reads
$$
\eqalign{
\ad_R (d_\q J_{R:X}) \cdot \q = \ad (X) \cdot \q, \forall X \in h_R, \cr
\ad_\s (d_\q J_{\s:X}) \cdot \q = \ad (X) \cdot \q, \forall X \in h_\s. \cr}
\eqn\generators
$$
Observe that this relates the gauge transformation in terms of
the $\Ad$-action
to the two $\Ad$-actions $\Ad_R$ and $\Ad_\s$ employed in the Poisson
brackets.

Consider the $\O_R$ orbit. We need the group $H_R$ to preserve the
orbit, in particular this implies that $H_R$ stabilises $\e$ such that
the term $\L_1=z \e$ is preserved.
Equation \generators\ has the form
$[\L, d_\q \hat J_{R:X}]=
[\L, X]$.
This implies that $d_\q \hat J_{R:X}=X$, which integrates to $\hat J_{R:X}=
\langle \q, X \rangle$. Comparing to equation \hamfunc, we obtain the
momentum map
$$
J_R : \hat g \rightarrow \hat g \big/ {\rm Ann}(h_R).
\eqn\quotmap
$$
The dual map $J^*_R : \C(\S,h_R) \rightarrow \C(\S,\hat g)$
is the inclusion map. The trivial calculation:
$$
\langle J_R(\Ad (g) \cdot \q), X\rangle
=\langle \q , {\rm Ad}(g^{-1}) \cdot X \rangle \equiv
\langle \Ad (g) \cdot J_R(\q), X\rangle
$$ proves that $J_R$ is $\Ad$-equivariant.
Thus we can reduce the orbit $\O_R$ by any subgroup $H_R$ that
stabilises $\e$.
This momentum map has previously been constructed in relation to the
$A_n$-KdV hierachies, [\DS,\Segal], these hierarchies being constructed as
Hamiltonian reductions of the orbits. Thus from the point of view of
the $G_R$-orbits this description of the KdV hierarchies  is well known.

For the orbit $\O_\s$,
the requirement that $H_\s$ preserves the orbit is highly restrictive.
Define
the subgroup $\tilde N_- \subset \C(\sphere, G)$ as the subgroup that
preserves the orbit
and $\tilde n_-$ as the corresponding Lie algebra.
Thus for all $\tilde X \in \tilde n_-$ there exists  an $X
\in \C(\sphere, g \mod
\Ker(\e))$ satisfying $\ad_\s(X) \cdot \q=\ad(\tilde X) \cdot \q, \
\forall \q \in \O_\s$, which implies that $\tilde n_-$ not only
annihilates $\e$, but lies in $\C(\sphere, n_-)$.
More specifically, $\tilde n_-$ consists of matrices of the form
$$
\pmatrix{0&0&0& \cdots &0&0 &0&0 \cr A&0&0& \cdots &0&0&0&0 \cr
B&A&0& \cdots &0&0 &0&0 \cr
C&B-A^\prime&A& \cdots &0&0&0&0 \cr \vdots& \vdots& \vdots& &
\vdots &\vdots& \vdots &\vdots \cr
&&& \cdots &A&0&0&0 \cr
&&& \cdots &B-(n-4)A^\prime&A &0&0 \cr
&&& \cdots &C-(n-4)B^\prime&B-(n-3)A^\prime&A&0 \cr}
\eqn\smallergauge
$$
where $A,B...$ are $C^\infty(S^1)$ functions, and $n$ is the dimension
of the matrix.
The dual $\tilde n_-^*$ is the strictly upper
triangular analogue of \smallergauge. Observe that the derivative
terms are only present if the representation has dimension $n \geq 4$.

The moment map of the $G_\s$-orbit is less trivial than that of the
$G_R$-orbit treated previously.
We are required to solve
the equation, \generators\
$$
\left[ \e, d_\q J_{\s:X} \right]= \left[ \L, X \right], \forall X \in
\tilde n_-.
$$
This splits into the following two parts. Since an element of the orbit
has the form $\q= \Lambda+ \left[\e, Y \right]$,  $Y \in \C(\sphere,
g \mod \Ker(\e))$,
we have
$$
\left[ \e, d_\q J_{\s:X} \right]= \left[ \partial_x +I, X \right]
+\left[ \e, \left[Y,X \right] \right], \forall X \in
\tilde n_-.
$$
Observe that there is a constant term, and a term linear in the
variable $Y$, implying that the Hamiltonian function $J_{\s:X}$
is quadratic in $Y$. It takes the form
$$
J_{\s:X}(\q(Y))=-\langle Y, \left[ \partial_x+I, X \right] \rangle
+ {1 \over 2} \langle \left[\e, Y \right], \left[Y,X \right] \rangle,
\eqn\momentsig$$
in terms of the variable $Y$
parametrising the orbit $\O_\s$. This equation is well defined given
that $Y \in \C(\sphere,g \mod \Ker(\e))$,
because of the presence of $\e$ in the
second term, and the fact that for $X \in \tilde n_-$ there exists
an $s_X \in \C(\sphere, g)$ satisfying
$ \left[ \partial_x +I, X \right]= \left[ \e, s_X \right]$.
Calculation of the functional derivative $d_\q \hat J_{\s :X}$
verifies that this is the desired
Hamiltonian function.
This is accomplished by using the observation that
all tangent vectors of $\O_\s$ have a form $r = \left[ \e, u\right], u
\in \C(\sphere,g \mod \Ker(\e))$, and thus
$$
\langle d_\q J_{\s:X}, r \rangle= -\langle \left[\e, d_\q J_{\s:X}
\right], u \rangle = \deriv J_{\s:X}(\q(Y+ \epsilon u)).
$$

{}From \momentsig\ we obtain
the momentum map $J_\s : \hat g^* \rightarrow \tilde n_-^*$,
$$
J_\s(\q(Y))=\left[\partial_x +I, Y\right]+{1 \over 2}
\left[ \left[\e, Y \right],Y \right]\  \mod \Ann(\tilde n_-).
\eqn\momsigY
$$

To perform the Hamiltonian reduction with respect to $J_\s$,
it is necessary to pull back a point $n_\s \in  \C(\S,\tilde n_-^*)$ and reduce
by the little group of $n_\s$. To calculate the little group
of $n_\s$ we need to know the equivariance relation of $J_\s$,
\ie we require the action $\Psi$ of $\tilde N_-$ on
$\tilde n_-^*$ such
that $J_\s(\Ad(g) \cdot \q)=\Psi(g) \cdot J_\s(\q), \forall g \in
\tilde N_-$.
We perform an infinitesimal calculation. Define the function $s :
\tilde n_- \rightarrow \C(\sphere,g \mod \Ker(\e))$ by
$ \left[ \partial_x +I, X \right]= \left[ \e, s_X \right]$.
Then we have
$$\eqalign{
\deriv J_\s(\q(Y + \epsilon(\left[Y,X \right]+ s_X)))=&
\left[ \partial_x +I, \left[ Y,X \right]+s_X \right]
+{1 \over 2} \left[ \left[\e, \left[Y,X \right]+s_X\right], Y\right] \cr
&+{1 \over 2} \left[ \left[\e, Y \right], \left[Y,X \right]+s_X\right]
\ \mod \Ann(\tilde n_-).\cr}
$$
This rearranges as follows
$$
\deriv J_\s\left(\Ad (e^{-\epsilon X}) \cdot\q(Y) \right)=
\left[ J_\s (\q(Y)), X \right]+\left[\partial_x+ I, s_X \right]\
\mod \Ann(\tilde n_-).
\eqn\eqvariancesig
$$
{}From this formulae it appears that the action on $\tilde n_-$ is
extended. Thus there exists a 1-cocycle $\sigma : \tilde N_-
\rightarrow \tilde n_-^*$ such that
the momentum map $J_\s$ is equivariant with respect to the action
$\Psi : (g, \eta) \rightarrow \Ad(g^{-1}) \cdot \eta + \sigma(g)$,
where $\sigma(g)= J(\Ad (g) \cdot \q)- \Ad (g) \cdot J(\q)$.
The 1-cocycle $ \sigma$ is the `exponential' of the additional term
in \eqvariancesig, and takes the form
$$
\sigma(g^{-1})= \left[\partial_x +I, S_g \right]
+{1 \over 2} \left[ \left[ \e , S_g \right], S_g \right] \mod \Ann(n_-),
\eqn\cocycle
$$
where $S$ is a map $S :\tilde N_- \rightarrow \C(\sphere,
g \mod \Ker(\e))$ defined by
$$
g^{-1} \left( \partial_x +I\right) g= I+ \left[ \e, S_g \right], \forall g
\in \tilde N_-.
\eqn\Sbig
$$
The proof is ommitted, being a specific case of theorem 3,
section 9.
The fact that $\s$ is a 1-cocycle follows from general Hamiltonian
reduction theory. By theorem 1, the
gauge algebra will be centrally extended
under Poisson brackets, with the cocycle (using \eqvariancesig )
$$
\Sigma(X,Y)= -\langle \e, [s_X, s_Y] \rangle, \forall X, Y \in \tilde n_-,
$$
which gives the following
Poisson bracket algebra
$$
\{ \hat J_{\s : X}, \hat J_{\s : Y} \}=
\hat J_{\s: [X,Y]}- \langle \e, [s_X, s_Y] \rangle,
$$
a central extension of $\tilde n_-$.

\chapter{THE GAUGE REDUCTION OF THE $G_R$-ORBIT}

In this section we consider the $G_R$-action on $C^\infty(\sphere,\hat
g^*)$ and
analyse the observation that the orbit through $\Lambda$
is only able to reproduce mKdV type potentials.
The only remaining degrees of freedom in the construction are the
${\bf s}[w]$-grade $\leq0$ components of the original point,
\ie $\q_0^{\leq0}$.
Through the choice of $\q^{\leq0}_0$ we
propose that the $G_R$-orbits split the KdV potentials, $\{u_k(x)\}$,
into various functional forms that are preserved under the flows of the
hierarchy.

Consider the traditional KdV hierarchy with $\Lambda$ of ${\bf
s}[w_{{\rm co}}]$-grade 1.
The $G_R$-orbit has the form
$$
n_-^{-1}b_+^{-1} \left(\partial_x+\q_0 \right) b_+ n_-+ z \e,
\eqn\mkdv
$$
where we have decomposed the action into two parts
$n_-(x) \in N_-, b_+(x) \in B_+$.
This orbit contains dynamical degrees of freedom that cannot be
reproduced by a $G_\s$-orbit, \orsig. In particular
the upper triangular dynamical components must be removed, which is
accomplished by an Hamiltonian reduction with respect to $\C(\sphere,
N_-)$. The appropriate value of the momentum map is
$n_R=I \in \C(\S,n^*_-)$, which gives the
level set constraint as
$$
P^{>0}_0 \left( b_+^{-1} \left(\partial_x+\q_0 \right)b_+ \right)=I.
\eqn\tradlevel
$$
The  action by $\C(\sphere,N_-)$ preserves this constraint, and is thus the
little group. We further observe that
the choice $n_-=1$ is a valid gauge choice, and hence
if the orbit \mkdv\  passes through $\Lambda$, ($\q_0=I$),
the theory is only capable of reproducing mKdV type potentials.
Since the orbits
never intersect, we further deduce that under the time evolution of
the hierarchy the potential always remains as an mKdV potential, a
familiar observation. To reproduce general KdV type potentials,
we propose that the components $\q^{\leq0}_0$ of the original point
are varied, \ie we consider the set of  orbits that foliate the space
$z \e +\C (\sphere, \hat g_{0} \cap \hat g^{\leq 1})$.
Because the
little group is identical for each orbit, the Hamiltonian reduction of
the $G_R$-orbits induces a map $\C(\sphere, \hat g_0) \rightarrow
{\cal M}$ that preserves the property of foliation, \ie leaves are
mapped to leaves.
This gives a construction for the
leaves of the foliation induced from the poisson bracket $\{\, ,\,
\}_R$, [\weinstein], as Hamiltonian reductions of $G_R$-orbits.
The fact that we are considering a foliation by reduced $G_R$-orbits
means that there is nothing more to prove, \ie
we only need to verify that a symmetry reduction exists such that
the level sets are submanifolds of
$\C (\sphere, \hat g_{>0} \cap \hat g^{\leq 1})$.
This implies that from the point of view of the Poisson bracket $\{\,
, \, \}_R$, the phase space is not $\cal M$, but a submanifold of
$\cal M$ given by the gauge equivalence classes of one of the  level sets.

An interesting question is the number of group orbits that foliate
the phase space $\cal M$.
This is of interest because for each
distinct orbit, we obtain a certain parametrisation of the potentials
$\{ u_k(x) \}$.
For example, as indicated above we obtain a distinction between mKdV
type potentials and {\it true} KdV type potentials depending on the
type of $G_R$-orbit from which the potential originates.
Thus,
each choice of orbit $\O_R$ containing a point $\q=\Lambda+\q^{\leq0}_0$
gives a set of potentials $\{u_k\}$ that {\it cannot} be described by
any other $G_R$-orbit.
In addition, since the flows of the hierarchies
are generated by the Poisson bracket $\{\, , \,\}_R$,
this partitioning of the
functional form of the potentials $\{u_k\}$ is also preserved by the
flows of the hierarchy.
The Muira maps $\{ {\cal M}_{{\bf s}}, \{\, , \, \}_{{\bf s}} \}
\rightarrow \{ {\cal M}_{{\rm co}}, \{\, , \, \}_2 \}$ are a weakened
form of this conclusion, {\it e.g.}\
at the crudest level this gives
the decomposition into mKdV type (passing through $\Lambda$), and non
mKdV type solution (from orbits that do not pass through $\Lambda$).
The exact nature of this partition is unknown.
In this paper we do not attempt to analyse this foliation of
the phase space $\cal M$ further.

We know from [\burr] that for gauge invariant functionals the Poisson
bracket $\{\, , \, \}_R$ is expressible in terms of the alternative
gradation endomorphism $R[w]$, $R[w]=P^{\geq0}-P^{<0}$.
This suggests that the phase space
can be analysed in terms of a $G_{R[w]}$-action.
Using the constraint \tradlevel, we observe that the orbit
\mkdv\ takes the form
$$
n_-^{-1}b_+^{-1} \left(\partial_x+\q_0 \right) b_+ n_-+ z \e
\equiv P^{\leq0}\left(n_-^{-1}b_+^{-1}
\left(\partial_x+\q_0 \right) b_+ n_-\right)+ \Lambda.
$$
The first term can be rewritten as
$$
n_-^{-1}P^{\leq0}\left(b_+^{-1}
\left(\partial_x+\q_0 \right) b_+\right) n_- +
P^{\leq0}\left(n_-^{-1} I n_-\right),
$$
using the constraint \tradlevel, and the decomposition $\hat g_0 =\hat
g^{>0}_0 +\hat g^{\leq0}_0$.
Hence, the level set can be rewritten in terms of a $G_{R[w]}$-action
$$
n_-^{-1}\left(
P^{\leq0} \left(b_+^{-1} \left( \partial_x +\q_0^{\leq0} \right)b_+
\right) + \Lambda\right) n_-
\eqn\levelsets
$$
where $b_+$ satisfies the differential equation \tradlevel.
By fixing the $n_-$ component of the group
action, equation \levelsets\ implies that each level set (with
momentum $J_R(\q)=I$),
can be gauge
fixed to be a subspace of a $G_{R[w]}$-orbit.
Thus, a $G_{R[w]}$-orbit of the form $P^{\leq0} \left(b_+^{-1} \L b_+
\right) + \Lambda$ is a partially gauge fixed Lax operator corresponding to a
collection of level sets, and thus after gauge fixing will correspond
to a collection of leaves of the $\{\, ,\,\}_R$-foliation.
This implies that the $G_{R[w]}$-orbits induce a partition of the potential
type that is coarser than that induced by the $G_R$-action, but still
finer (or equal) to that into mKdV, pmKdV and {\it true} KdV type.

\chapter{THE CASE OF $Sl(2)$}

In this example, we explicitly construct the orbits in
$\C(\sphere,\hat {sl}(2)^*)$
under the two group actions $G_R$ and $G_\s$, exploiting the fundamental
representation of $Sl(2)$.
We perform the Hamiltonian
reductions with respect to the symmetry groups
$H_{R/ \s}=\C(\S,N_-)$,
and verify that for a certain choice of inverse images $J^{-1}_{R/
\s}$ the phase spaces are dynamically identical.

\vskip 4pt
\centerline{
{\it The $G_R$-orbit.}}

We split the action of $G_R$ into an action by $N_-$ and $B_+$. We
are interested in orbits that intersect the space $\q=z \e
+ X, X \in \C(\sphere, \hat g_0)$, with the upper triangular component
non-zero for all $x \in \sphere$. For these elements, we can use the
lower Borel subgroup, $B_-$, to express them in the form
$\Lambda +\mu \e, \, \mu \in \C(\sphere)$. Thus the
orbits of interest have the form :
$$\eqalign{
\q=& z \e+ b_+^{-1} \left( \partial_x + I +\mu \e\right) b_+ \cr & \cr
=& z\e+\pmatrix{\a^{-1} \a^\prime- \mu \a \b & \a^{-2} \left( \left(\a
\b\right)^\prime+1 \right)-\mu \b^2 \cr
\mu \a^2  & -\a^{-1} \a^\prime +\mu \a \b \cr}
{\rm \ for\ } b_+= \pmatrix{\a & \b \cr 0 & \a^{-1} \cr},\cr}
\eqn\orbitR
$$
with the additional $\C(\sphere,N_-)$
component of the group
action  suppressed.
The momentum map $J_R : \O_R \rightarrow n_-^* \cong n_+$ is
$$
J_R(\q)= \a^{-2} \left( \left(\a \b \right)^\prime +1 \right)
-\mu \b^2
\eqn\momentR
$$
Since $n_+$ is one dimensional we are suppressing the basis element in
this, and all following formulae.
To perform the Hamiltonian reduction, we restrict $\O_R$ to the
inverse image of a point $n_R \in \C(\S,n_-^*)$, and reduce by the little
group of $n_R$. For all choices of $n_R$ the little group is $\C(\S,n_-)$,
as follows from the abelian nature of $n_-$ and the fact that $J_R$ is
$\Ad$-equivariant. This
Hamiltonian reduction has been  considered before in [\DS,\Segal].

\vskip 4pt
\centerline{{\it The $G_\s$-orbit.}}

Under the action of $G_\s$ the orbit through $\Lambda$ takes the form:
$$
\q=\Lambda+ \left[ \e, Y \right]=
z\e+ \pmatrix{-B& 1 \cr 2A & B \cr}
{\rm \ for\ } Y=\pmatrix{A & B \cr * & -A \cr} \in C^\infty(\S,sl(2)).
\eqn\orbitsig$$
The momentum map $J_\s$ of the $\Ad$-action of $\C(\S,N_-)$ on this orbit takes
the form, \momsigY\
$$
J_\s(\q(Y))=B^\prime-B^2-2A.
\eqn\momentsig
$$
To carry out the Hamiltonian reduction by $\C(\S,N_-)$, we must
evaluate the function $s_X$
defined by $\left[ \partial_x +I, X \right]= \left[ \e, s_X \right],
\ X \in C^\infty(\sphere,n_-)$. It has the form
$$
s_X= \pmatrix{ {1 \over 2}C^\prime & -C \cr * & -{1 \over 2}C^\prime \cr},
{\rm \ for \ } X= \pmatrix{0 & 0 \cr C & 0 \cr}.
$$
This gives the following central extension to the Poisson bracket
algebra
$$
\Sigma(X,Y)= \int dx \langle \left[ \partial_x+I, X \right], s_Y
\rangle =\int dx \, {\rm tr} \left( \pmatrix{D & 0 \cr D^\prime & -D}
\pmatrix{ {1 \over 2} C^\prime & -C \cr * & -{1 \over 2} C^\prime}\right)
=2 \int dx\, D C^\prime ,
$$
for $X=D \e, Y= C \e \in C^\infty(\S, n_-)$.
Thus the Poisson bracket algebra is
$$
\{ \hat J_{\s: X}, \hat J_{\s: Y} \}_\s= 2 \int dx\, D C^\prime,
$$
since $n_-$ is abelian.
The appropriate action of $\C(\S,N_-)$ on $\C(\S,n_+)$ is given by
$$
\Psi(g) \cdot n_\s=n_\s + \sigma(g), \forall g \in N_-,
$$
where the cocycle $\sigma$ is defined in \cocycle.
The function $S_g$ has the form
$$
S_g= \pmatrix{{1 \over 2} \left(\c^\prime- \c^2 \right)& -\c \cr
*& -{1 \over 2} \left(\c^\prime- \c^2 \right) \cr},
{\rm \ for \ } g=\pmatrix{1 & 0\cr \c & 1 \cr},
$$
giving the cocycle $\sigma(g)=2\c^\prime$.
Hence
the little group of a point $n_\s \in \C(\S,n_-^*)$
only consists of constant elements of the gauge group, \ie
$G_\s(n_\s) = N_-$.

\vskip 4pt
\centerline{{\it A Comparison of the orbits.}}

Comparing the two orbits in \orbitR\ and \orbitsig, we observe that
they can never be identical because of the existence of a
dynamical degree of freedom in the upper triangular component of the
$G_R$-orbit. We conclude that an Hamiltonian reduction of the orbits
is necessary.
Recall that the first step in the Hamiltonian reduction is to specify
the constants of motion, \ie
choosing the points $n_R, n_\s \in \C(\S,n_-^*)$.
Comparing the two orbits \orbitR\ and
\orbitsig, we observe that the appropriate choice is
$n_R=1$ with the associated level set consisting of matrices
that satisfy
$$
\left(\a \b \right)^\prime +\mu \left( \a \b \right)^2+1 =\a^2.
\eqn\diff
$$
This is a differential equation for the combination $\d=\a \b$, given
$\a$. Reversing the logic, this equation gives $\a$ in terms of $\d$.
Thus the orbit has one
degree of freedom, $\d$.
The Lax operator now takes the form
$$
z\e+\pmatrix{\a^{-1} \a^\prime- \mu \d & 1 \cr
\mu \a^2  & -\a^{-1} \a^\prime +\mu \d \cr},
\eqn\orbitco
$$
where $\a^2$ is related to $\d$ through \diff. The gauge group introduces the
following equivalence relation
$$
\pmatrix{F &1 \cr G & -F \cr} \sim \pmatrix{F+\c & 1 \cr
G+\c^\prime -\c^2-2 \c F & -F-\c \cr}
\eqn\gaugeequiv
$$
the matrix expansion of the formula $\L \sim g^{-1} \L g,
\forall g \in C^\infty(\S, N_-)$.
As is well known, we can choose a gauge slice with no diagonal
component, which implies that there is only one dynamical field.
By the choice $\c=-F=-\a^{-1} \a^\prime+ \mu \d$,
we obtain the traditional gauge
slice with $u(x)=-F^\prime+F^2+\mu \a^2$. Thus the potential $u(x)$
has a rather complex functional form in terms of the dynamical
variable $\d$. Recall that under the flows of the hierarchy, this
functional form is preserved, since the $G_R$-orbit is preserved by
the flow. Due to the foliation of $\C(\sphere, \hat g^*)$ by the
orbits of $G_R$, the full space $\Lambda+\C(\sphere, \hat g_0 \cap
\hat g^{<1})$ is covered by the
level sets, and hence the full phase space $\cal M$ is reproduced, \ie
generic $u(x)$ can be reproduced through changes in $\mu(x)$.
Unfortunately, the number of orbits required is unknown, and the
corresponding classification of these orbits by choices for
the function $\mu \in \C(\sphere)$ is also unknown.

We performed an Hamiltonian reduction on the $G_R$-orbits
because they contained additional degrees of freedom over and above
those contained in the $G_\s$-orbit. This produced a system with only
1 dynamical field. In contrast, the orbit $\O_\s$ has 2 dynamical
fields, \ie $A$ and $B$, \orbitsig. Thus we must reduce the degrees of
freedom by 1. This is accomplished through an Hamiltonian reduction by
$J_\s$.
Imposing the level set constraint $J_\s(\q)=n_\s$, \momentsig, we obtain the
level set as
$$
\q=z\e + \pmatrix{-B &1 \cr B^\prime -B^2- n_\s & B \cr},
\eqn\mason
$$
which only passes through $\Lambda$ if $n_\s=0$.
Under equivalence with the little group $G_\s(n_\s) =N_-$,
we obtain the equivalence relation
$$
\pmatrix{-B &1 \cr B^\prime -B^2- n_\s & B \cr}
\sim \pmatrix{\c-B & 1 \cr \left(B+\c \right)^\prime -
\left(B-\c \right)^2-n_\s & B-\c \cr},
\eqn\primitive
$$
where  $\c$ is a constant.
This implies that the constant component of $B$ is gauge, and thus
non-dynamical. We deduce that
\mason\ should be a gauge slice for the
$sl(2)$-KdV hierarchy with the gauge fixing constraint $\int B(x) dx=0$.
This is proved by using the full gauge group $\C(\sphere, N_-)$, \ie
$\c$ arbitrary, to map
\mason\ into the traditional gauge slice with no Cartan subalgebra
component.
We derive
the relationship between the traditional potential $u(x)$ and
the dynamical field $B(x)$ as
$u(x) = 2 B^\prime- n_\s$, a relationship that supports the conclusion
that the constant component of $B$ is non-dynamical.
This parametrisation of $u(x)$ also
implies that the constant component $\int u(x) dx$ is not dynamical, \ie
a constant of motion. This is in fact observed in the traditional
analysis of the KdV hierarchy, since $\int u(x)dx $ is the Hamiltonian
that generates the chiral flow, and is thus constant under all the
flows.
We conclude that \mason\ is a good gauge slice for the theory, a
gauge slice that was in fact used in [\MS], in the derivation of the
$sl(2)$-KdV hierarchy from the Self Dual Yang-Mills Equations. We shall
find that in our
other examples, we also obtain a gauge fixing that is similar to
this gauge slice, and refer to them as Mason-Sparling gauge slices.

We deduce that a coadjoint orbit construction with group action
$G_\s$ is able to reproduce the phase space of the $sl(2)$-KdV
hierarchy. This provides an underlying group theoretic
description  for
the first Poisson bracket $\{\, , \,\}_1$.
In addition, the leaves of the foliation of
the phase space $\cal M$ defined by the second poisson bracket $\{\,
,\,\}_R$ are gauge group reductions of $G_R$-orbits.

\chapter{THE CASE OF $Sl(3)$}

In [\burr, \Tim], three theories were constructed from the Kac Moody
algebra $sl(3)$, with the element $\Lambda$ taking  the various
regular forms
$$
\Lambda_{{\rm co}}=\pmatrix{0&1&0 \cr 0&0&1 \cr z&0&0 \cr},\
\Lambda^2_{{\rm co}}=\pmatrix{0&0&1 \cr z&0&0 \cr 0&z&0 \cr},\
\Lambda_{\alpha}=\pmatrix{0&0&1 \cr 0&0&0 \cr z&0&0}
$$
Recall that the $G_\s$-orbit only depends on the
components of $\L$ with homogeneous degree greater than zero, \ie
$\L_{>0}$.
Thus the $G_\s$-orbits of  $\Lambda= \Lambda_{{\rm co}}$
and $\Lambda=\Lambda_\alpha$
are essentially identical,
\ie the orbits are linear translates of each other. The
difference in these theories lies in their different
gauge group actions, leading to different dynamical degrees of freedom.
In this section we perform an  analysis of the case $\Lambda=
\Lambda_{{\rm co}}$, the
traditional $sl(3)$-KdV hierarchy. The remaining two theories are
examined in later sections, and serve as useful comparisons.
The traditional hierarchy, $\Lambda=\Lambda_{{\rm co}}$,
is constructed from the Lax operator
$$
\L=\partial_x+
z\e+ \pmatrix{U^+ & 1 &0 \cr  G^+ & -U^+-U^- &1 \cr T &  G^- & U^- \cr},
\eqn\thrpot
$$
with an equivalence relation
$\L \cong n_-^{-1} \L n_-, \forall n_- \in \C(\sphere, N_-)$.
This implies that \thrpot\ is equivalent to
$$\eqalign{
&n_-^{-1} \L_0 n_-=\cr \cr
&
\pmatrix{U^++\a & 1 &0 \cr
G^++\a^\prime+\c-\a\left(2U^++U^-+\a\right) & -U^+-U^--\a+\b &1 \cr
\hat T&
G^-+\b^\prime-\c+\b\left(U^++2U^-+\a-\b\right) &U^--\b\cr},\cr
\cr
& \ \ \ \ \ \ \ \ \ \ \
{\rm \ with\ \ }
\matrix{\cr \hat T=T+\c^\prime-\b\a^\prime+G^-\a-G^+\b+\a \b(2U^++U^-) \cr
\ \ \ \ \ \ \ \ \ \ +\a^2\b-(U^++\a-U^-+\b)\c \cr},\
n_-=\pmatrix{1 &0&0\cr \a & 1&0 \cr \c & \b & 1 \cr}. \cr}
\eqn\thrgauge
$$
The gauge degrees of freedom in
$\a,\b,\c$ are able to eliminate the matrix elements $U^\pm,G^+$
respectively, reproducing the traditional gauge slice.

The above Lax operator can be reproduced from an $\O_\s$-orbit.
The orbit $\O_\s$ takes the form
$$
\q=\hat \q+\left[\e, Y\right]=
\hat \q+\pmatrix{-B&0&0\cr -C&0&0 \cr 2D &A &B\cr},
\ {\rm for\ } Y=\pmatrix{D&A&B\cr *&*&C \cr *&*& -D\cr}\in
C^\infty(\sphere,sl(3)),
\eqn\Rorbits
$$
where $\hat \q \in \C(\sphere, \hat g_{\geq0} \cap \hat g^{\leq 1})$
is the original point of the orbit, satisfying the restriction
$P^{\geq1} \left(\hat \q \right)=\Lambda$.
This orbit has four degrees of freedom, too many to be equivalent to the
$sl(3)$-KdV hierarchy. We reduce the number of degrees of freedom by
two by an Hamiltonian reduction with $\tilde N_-$, the subgroup of
$\C(\sphere, N_-)$ that preserves the orbit.
With $\Lambda=\Lambda_{{\rm co}}$, the
momentum map $J_\s$ for the $\Ad(\tilde N_-)$-action reads
$$
J_\s(\q(Y))=\left[\partial_x+I,Y \right]+{1 \over 2}\left[\left[\e,
Y\right],Y \right]=\pmatrix{*&\hat A^\prime -D -B \hat A &
B^\prime -A +C -B^2 \cr *&*& \hat A^\prime -D -B \hat A \cr
*&*&* \cr},
$$
where we realise $C^\infty(\sphere, n^*_-)$ as the matrix analogue
of \smallergauge, and define $\hat A={1 \over 2}(A+C)$. Thus the level set
with momentum   $J_\s(\q)= -{1 \over 2}\mu I- \nu e_{\psi}$
has the form
$$
\q= \Lambda_{{\rm co}}+
\pmatrix{ -B&0&0 \cr
-\hat A+ {1 \over 2}\left(B^\prime -B^2 +\nu\right) &0&0\cr
\mu+ 2\hat A^\prime -2B \hat A&
\hat A+ {1 \over 2}\left(B^\prime -B^2 +\nu\right) & B \cr},
\eqn\msgauge
$$
and passes through the point
$\Lambda+\mu \e +{1 \over 2}\nu (e_{-1}+e_{-2})$.

As in the case of $sl(2)$, the little group of a
point $n_\s \in \C(\S, \tilde n_-^*)$ is not the full gauge group,
$G_\s ( n_\s)
=\{g \in \C(\S, \tilde N_-)\, | \, \sigma(g)=0 \}$ since $\tilde N_-$
is abelian.
Using definition \Sbig, we find that
$$
S_g=\pmatrix{{1 \over 2} \left(\c^\prime- \a
\left(2 \c+\a^\prime -\a^2 \right) \right)
& -\c+\a^\prime & -\a \cr
* & * & -\c-\a^\prime+\a^2 \cr
*& *& -{1 \over 2} \left(\c^\prime-\a
\left(2 \c+\a^\prime -\a^2 \right)\right)},
$$
which implies that the cocycle has the form
$$
\sigma(g^{-1})= \pmatrix{*& -{3 \over 2} \left(\c^\prime -\a \a^\prime
\right) & -3 \a^\prime \cr *&*& -{3 \over 2} \left(\c^\prime -\a
\a^\prime \right) \cr *&*&* \cr},{\rm \ where\ }
g =\pmatrix{1 &0 &0 \cr \a & 1 & 0 \cr \c & \a & 1}.
$$
The little group is given by elements $g \in \tilde N_-$
satisfying the constraint $\sigma(g)=0$, \ie $\a^\prime=\c^\prime=0$.
Thus the little group is
the constant subgroup of $\tilde N_-$.
Under equivalence by $\Ad (G_\s (n_\s))$ we deduce that the constant
components of $\hat A$ and $B$ are non-dynamical, the
little group defining the equivalence $\hat A \sim \hat A-\c+{1\over
2} \a^2, B \sim B-\a$, $\a, \c$ arbitrary constants, \thrgauge.
This equivalence can be fixed by imposing
$\int \hat A dx=\int B dx=0$. With these constraints,
\msgauge\ should be a gauge slice of the $sl(3)$-KdV hierarchy,
a gauge slice that is
the $sl(3)$ analogue of the Mason-Sparling gauge slice, [\MS].
This is proved by showing that the gauge slice \msgauge\
is equivalent to that traditionally employed. This
involves the construction of a gauge transformation that transforms
\msgauge\ into the traditional gauge slice, such that the the
parametrisations are in 1:1
correspondence.
Using full gauge equivalence, we obtain the relations
$u_1(x)=\mu+3 \hat A^\prime-{3 \over 2} B^{\prime \prime},
u_2(x)=3 B^\prime+\nu$. These expressions confirm
the conclusion that the
constant components of $\hat A, B$ are non-dynamical.
Thus the gauge slice \msgauge\ is equivalent
to the traditional slice of [\DS]
under full gauge equivalence since generic potentials $u_1(x), u_2(x)$
can be reproduced. The only restriction is that the Fourier components
$\int u_1 dx, \int u_2 dx $ are non-dynamical. These are Hamiltonians
as in the case of $sl(2)$, confirming that they are constants of motion.

We have proved that the traditional $sl(3)$-KdV
hierarchy can be constructed as an
Hamiltonian reduction of a $G_\s$-orbit. We now wish to prove that the
$G_R$-orbits are able to reproduce the dynamics of this system, \ie
that the orbits $\O_R$ have an Hamiltonian reduction
such that the level sets
possess a gauge fixing to  the traditional gauge slice, or
equivalently to \msgauge.
The $G_R$-orbits have the form $g^{-1} \L_0 g +\L_1$, and thus foliate
the space $\C(\sphere, \hat g_0) +\L_1$. Since $\C(\sphere, N_-)$ stabilises
$\L_1=z \e$, we can Hamiltonian reduce with respect to
$\C(\sphere, N_-)$. Comparing to  the $G_\s$-orbit,
the appropriate momentum is
$J_R(\q)=I$. The corresponding little group is $\C(\sphere, N_-)$, the
full gauge group.
Thus the level sets are of the form \thrpot\ with a gauge equivalence
under $\C(\sphere, N_-)$. Hence the reduced phase spaces are the symplectic
leaves of $\cal M$ as required, proving that the $sl(3)$-KdV hierarchy
can be described dynamically as reduced $G_R$-orbits.
This foliation appears to be  finer than the split of the
potentials into mKdV, pmKdV, and {\it true} KdV type potentials. For
example, orbits that possess a point with the form
$$
\pmatrix{0&1&0 \cr \nu &0&1 \cr 0& \nu &0 \cr}
\eqn\pKdV
$$
can only produced pmKdV (or mKdV) type potentials, since only through an action
by the gauge group $\C(\sphere, N_-)$ can a term proportional to $\e$
be reproduced. By definition, this is the pmKdV type potential, [\Tim].
We note however that the pmKdV type potentials are probably not
covered by only one $G_R$-orbit, since there exist other initial
points
than the choice \pKdV\ that do not lie on the $G_R$-orbit of \pKdV.

The alternative theory with $\Lambda=\Lambda_{{\alpha}}$
can also be analysed in terms of the orbit structures.
It contrasts very nicely with the theory associated with
$\Lambda=\Lambda_{{\rm co}}$, showing that
a different choice of $\Lambda_0$ exist such that the reduced
orbits describe bi-Hamiltonian systems.
This theory is pursued further in section 11 after we have extended the
analysis of sections 5 and 6 to include this more general $\Lambda$.

\chapter{THE GAUGE GROUP MOMENTUM MAPS AND HAMILTONIAN REDUCTION II}

In this section we extend the momentum map analysis of
section 5 to include theories constructed with $\Lambda$ of
${\bf s}[w]$-grade greater than 1, \ie with a general form $\Lambda=
zI_- +I_+$.

Define the homogeneous decomposition of
$\Lambda$ as $\Lambda=zI_-+I_+$, where $I_\pm$
only have components of
${\bf s}[w]$-grade $\leq r_\pm< N$
respectively, $r_+=N+r_-=i>0$.
We have not given $\Lambda$ a definite ${\bf s}[w]$-gradation as
suggested in [\Tim]. However this structure will be reproduced after a
study of the dynamics, when we separate the components
in $\Lambda$ into those with ${\bf
s}[w]$-grade $<r_+$ as constants of motion (or absorbed into dynamical
fields), and  identify
the components with ${\bf s}[w]$-grade $r_+$ as the element of the
Heisenberg algebra ${\cal H}[w]$ called $\Lambda$ in [\Tim].
The $G_\s$-orbit takes the form
$$
\q=\Lambda +\left[I_-,Y \right],\  Y \in \hat g_0 \mod
\Ker(I_-).
\eqn\Iorbit
$$
Upper and lower subscripts denote principal and homogeneous gradation
respectively.
As in section 5, the space
$\C(\sphere,\hat g_0 \mod \Ker(I_-))$ is used to parametrise the orbit $\O_\s$.

Define the gauge group $\tilde N_-$ as the subgroup of
$C^\infty(\sphere, N_-)$ that preserves the orbit \Iorbit, and $\tilde
n_-$ the corresponding Lie algebra.
This implies that $\tilde n_- \subset \Ann(I_-)$, since
if $\left[I_-, X \right] \not=0$ the
gauge group action would not preserve the homogeneous grade 1
term, \ie $z I_-$.
We  make the following definition

\df
{\it Define the map $s : \tilde n_- \rightarrow \C(\sphere,\hat g_0 \mod
\Ker(I_-))$ as}
$$
\left[ \partial_x +I_+,X \right]= \left[ I_-, s_X \right],
$$
{\it such that under infinitesimal gauge transformations we have the
relation}
$$
\left[ \partial_x+\q, X \right]=\left[I_-, s_X +\left[Y,X
\right]\right], \forall X \in \tilde n_-.
$$

Given a symmetry $\Phi : G \times P \rightarrow P$, the
momentum map is calculated by solving equation \momorbit,
which takes the form
$$
\ad_\s (d_\q J_{\s:X}) \cdot \q = \ad (X) \cdot \q, \forall X \in
\tilde n_-.
$$
This has the solution
$$
J_{\s:X}(\q(Y))=-\langle Y, \left[ \partial_x+I_+, X \right] \rangle
+ {1 \over 2} \langle \left[I_-, Y \right], \left[Y,X \right] \rangle,
\eqn\momentsigtwo
$$
in terms of the variable $Y \in \C(\sphere, g \mod \Ker(I_-))$, that
parametrises the orbit $\O_\s$.
This is verified through a calculation
of the functional derivative $d_\q \hat J_{\s :X}$,
accomplished by using the observation that
all tangent vectors of $\O_\s$ have a form $r = \left[ I_-, u\right], u
\in \C(\sphere,g \mod \Ker(I_-))$. Thus
$$
\langle d_\q J_{\s:X}, r \rangle= -\langle \left[I_-, d_\q J_{\s:X}
\right], u \rangle = \deriv J_{\s:X}(\q(Y+ \epsilon u)).
$$

{}From \momentsigtwo\ we obtain
the momentum map $J_\s : \hat g^* \rightarrow \tilde n_-^*$,
$$
J_\s(\q(Y))=\left[\partial_x +I_+ , Y\right]+{1 \over 2}
\left[ \left[I_-, Y \right],Y \right]\  \mod \Ann(\tilde n_-).
\eqn\momsigYtwo
$$

The Hamiltonian reduction with respect to $J_\s$ is accomplished
by restricting to the pull back of a point
$n_\s \in \tilde n_-^*$ and reducing
by the little group of $n_\s$. To calculate the little group
of $n_\s$ we need to know the equivariance relation of $J_\s$,
\ie we require the action $\Psi$ of $\tilde N_-$ on
$\tilde n_-^*$ such
that $J_\s(\Ad(g) \cdot \q)=\Psi(g) \cdot J_\s(\q), \forall g \in
\tilde N_-$. Following the calculation in section 5, we define the
`exponential' of the map $s$

\df
{\it Define the map $S :\tilde N_- \rightarrow \C(\sphere,g \mod
\Ker(I_-))$ by}
$$
g^{-1} \left( \partial_x +I_+ \right) g=
I_+ + \left[ I_-, S_g \right], \forall g
\in \tilde N_-.
$$

\thm
{\it The equivariance relation of the gauge group momentum map
is with respect to the action
$\Psi : (g, \eta) \rightarrow \Ad(g) \cdot \eta + \sigma(g)$,
where $\sigma(g)$ is a 1-cocycle of
$\tilde N_-$ given by}
$$
\sigma(g^{-1})= \left[\partial_x +I_+, S_g \right]
+ {1 \over 2} \left[ \left[I_-, S_g \right], S_g \right] \mod
\Ann(\tilde n_-).
\eqn\cocycletwo
$$

\prf
This is proved by calculating $J_\s \left(\Ad(g^{-1} ) \cdot \q
\right)$ and extracting the $\Ad$ action by $g \in G$. Using
definition 5, we obtain
$$\eqalign{
& J_\s \left(\Ad(g^{-1} ) \cdot \q(Y) \right) \equiv
J_\s \left( \q \left( \A (g^{-1}) \cdot Y +S_g \right) \right)= \cr
&\ \ \
\left[ \partial_x +I_+, \A (g^{-1}) \cdot Y +S_g \right]
+{1 \over 2} \left[\left[I_-, \A(g^{-1}) \cdot Y +S_g \right],
\A(g^{-1}) \cdot Y +S_g \right] \cr}
$$
We now rearrange, the second term taking the form
$$
{1 \over 2} \A (g^{-1}) \cdot \left[ \left[I_-, Y \right], Y \right]
+{1 \over 2}\left[ \left[ I_-, S_g\right], S_g \right]
+\left[ \left[I_-, S_g \right], \A (g^{-1}) \cdot Y\right],
$$
where terms of the form
$[I_-, \, ]$ are zero after taking the quotient with $\Ann(I_-)$.
The cocycle \cocycletwo\ now follows. The fact that this is a cocycle
can be proved by using the relation $S_{gh}=h^{-1}S_g h +S_h$.

By theorem 1, the
gauge algebra will be centrally extended
under Poisson brackets,
$$
\{ \hat J_{\s : X}, \hat J_{\s : Y} \}_\s =
\hat J_{\s: [X,Y]}- \langle I_-, [s_X, s_Y] \rangle,
$$
a central extension of $\tilde n_-$.

\chapter{THE GAUGE REDUCTION OF THE $G_R$-ORBIT II}

In this section we generalise the
analysis of section 6 to include the case of a more general $\Lambda$,
$\Lambda=I_+ +zI_-$. We assume that there exists a gradation ${\bf s}^\prime$
such that the space $\oplus_{k=1}^{i-1} \hat g_k({\bf s}^\prime)
\cap \hat g_{\geq 0}({\bf s})$ lies
in the image of $\A (P_-) \cdot \Lambda -\Lambda$, where
$P_-$ is the Lie group generated by
$\hat g_0({\bf s}) \cap \hat g_{<0}({\bf s}^\prime)$.
We further assume
that the gradation ${\bf s}^\prime$
is maximal with respect to $\preceq$ for all
gradations satisfying this property, and that $s_0^\prime \not=0$.
In [\Tim], the gradation ${\bf s}^\prime$ is required to be induced
from an equivalence class $[w]$ of the Weyl group. This will also be
assumed,  ${\bf s}^\prime ={\bf s}[w]$,
but takes no part in the following calculations.

With $\Lambda=I_+ +z I_-$, $I_\pm$ having components
of ${\bf s}[w]$-grade $\leq r_\pm$ respectively,
the $G_R$-orbit takes the form
$$
\q=n_-^{-1} b_+^{-1} \left( \partial_x  +I_+ \right) b_+ n_-
+z I_-
\eqn\genorbit
$$
where we choose $\q=I_+$
for the original point. The fact that $I_+$ has negative
${\bf s}[w]$-grade components implies that \genorbit\ describes all
orbits of this type.
As in section 9, the fact that $\Lambda$ has a well defined ${\bf
s}[w]$-grade will be reproduced in the final analysis.
Note that only degrees of freedom
generated through the action of $G_R$ are dynamical. In particular
this means $\L_1$ is non-dynamical.
Thus we must allow $I_+$ to contain components with ${\bf s}[w]$-grade
less than $r_-=r_+-N$ in order to
achieve exact equality with the KdV hierarchies of [\Tim].

Recall that
the generalised KdV hierarchies are defined on the phase space $\cal
M$, gauge equivalence classes of $\C(\sphere, \hat g_{\geq0} \cap \hat
g^{<i})$ with the gauge group generated by $\C(\sphere, \hat g_0 \cap \hat
g^{<0})$. This suggests that we attempt to Hamiltonian reduce the
$G_R$-orbits with respect to the group $\C(\sphere, P_-)$.
However, under $\Ad$-action this group does not preserve
the orbits \genorbit\ for general $I_-$.
We now have two courses of action, use the smaller
gauge group preserving the orbits, or increase the space we wish
to reduce to the image of the $\C(\sphere, P_-)$
action. The second option removes us from the theory of Hamiltonian
reduction, the image only having a Poisson structure.
Pursuing the first option,
define the symmetry group $\tilde N_- \subset \C(\sphere, P_-)$ as the
subgroup that stabilises $I_-$ under $\A$-action,
\ie $\A (\tilde N_-) \cdot I_-=I_-$.
Under $\Ad$-action, $\tilde N_-$ preserves the orbits \genorbit, and
has a momentum map $J_R : \O_R \rightarrow \C(\sphere,g^*
\mod \Ann(\tilde n_-))$. Performing an Hamiltonian reduction with
respect to $\tilde N_-$, we choose the level set constraint
$$
P^{>0} \left( n_-^{-1} b_+^{-1} \left( \partial_x +
I_+ \right) b_+ n_- \right) = I_+ \mod \Ann(\tilde n_-)
\eqn\genlevel
$$
where we have assumed with no loss of generality that $I_+$ is the
value of the momentum map, \ie we generate the orbit from  a point on
the level set of interest.
The assumption that $s_0^\prime \not=0$ implies that all lowering
operators of ${\bf s}[w]$-gradation $\leq -r_+$ are members of the Lie algebra
$\tilde n_-$. This implies that all the components of the Lax
operator with ${\bf s}[w]$-grade $\geq r_+$ are
totally specified by \genlevel.

The little group $G(I_+)$ is not in general the full symmetry group
$\tilde N_-$.
However it is identical for each orbit. Thus the Lax operator has a
form
$$
\L= \partial_x + \Lambda + q(x),\, q(x) \in \Ann(\tilde n_-) \cap \hat
g_0,
\eqn\pgauged
$$
with a gauge equivalence under the group $G(I_+) \subset \tilde N_-$.
The fact that the little group is not $\tilde N_-$ implies that the
Lax operator
is partially gauge fixed, \ie we can increase the equivalence relation
to an equivalence under $\tilde N_-$ invoking
the penalty of altering the momentum. Only  the momentum with
${\bf s}[w]$-grade $<r_+$ can be altered because $\tilde n_- \in
\C(\sphere, \hat g_0 \cap \hat g^{<0})$, and the assumption
regarding the image of $\A (P_-)
\cdot \Lambda$ implies that all momenta of this type are altered.
We
deduce that the space of Lax operators of the form \pgauged\ is
equivalent to the space of Lax operators with the form
$$
\L=\partial_x +\Lambda +q(x),\, q(x) \in \hat g_0 \cap \hat g^{<i},
$$
with gauge group $\tilde N_-$. This is similar to the structure in
[\Tim], except for the treatment of the homogeneous grade 1
components, and smaller gauge group.
The question arises as to whether this is a partial gauge fixed form of
a generalised KdV Lax operator, the partial gauge fixing corresponding to the
reduction $\C(\sphere, P_-) \rightarrow \tilde N_-$.
Since the only constraint on $\tilde N_-$ is that it stabilises $I_-$,
we can increase the gauge group to $P_-$ if we treat the components of
${\bf s}[w]$-grade $<r_-$ in $I_-$ as dynamical.
Thus we extract the fact that $\Lambda$ has well defined ${\bf s}[w]$
through an increase of the symmetry group from $G(I_+)$ to
$\C(\sphere, P_-)$.

Since we have enlarged the equivalence relation in \pgauged\ to
an equivalence under $\tilde N_-$, the momentum constraint
becomes
$$
P^{\geq i} \left(  b_+^{-1} \left( \partial_x +
\q_0 \right) b_+  \right) = P^{\geq i} I_+.
\eqn\constraint
$$
Thus the level set of the orbit \genorbit\ can be rewritten as
$$
\q=n_-^{-1} P^{\leq0} \left(b_+^{-1} \left( \partial_x +\q_0 \right)
b_+ \right) n_-
+ n_-^{-1} I_+ n_-
+z I_-.
$$
{}From the definition of  $\tilde N_-$, we have $n_-^{-1} I_- n_-=I_-$,
and hence the level set is gauge equivalent to a subspace of a
$G_{R[w]}$-orbit. Thus we again obtain the conclusion that the
$G_{R[w]}$-orbits induce a partition of the potentials that is coarser
than that defined by the $G_R$-orbits.

Finally in this section, we reexamine the reduction of the $G_R$-orbits
by $\C(\sphere, P_-)$. Since the $G_R$-orbits are not preserved under the
$\Ad$-action by $\C(\sphere, P_-)$, we consider the space
$$
\{ g^{-1} \L_0 g + zI_- + \hat u \, \big|\, g \in G, \hat u \in
\C(\sphere, \hat g_1 \cap \hat g^{<i} )\},
\eqn\poissonman
$$
where $\L_0= \partial_x+I_+ + q$, with $q$ a point in $\C(\sphere,
\hat g_0 \cap \hat g^{<i})$. We take $I_+, I_-$ as having well defined
${\bf s}[w]$-grade, such that
$\Lambda=I_+ +z I_-$ has ${\bf s}[w]$-grade $i$.
This is a Poisson manifold,  Poisson bracket $\{\, , \,\}_R$, and
kernel  generated by the coordinate $\hat u$.
To perform an analogue of Hamiltonian reduction, we consider the space
of equivalence classes of \poissonman\ under $\Ad (\C(\sphere,P_-))$,
and construct a leaf
of the foliation induced by the Poisson bracket $\{\, ,\, \}_R$, [\weinstein].
Using the corresponding Hamiltonian reduction of $\O_R$ as a guide, we
construct the gauge invariant functions $\hat J_X$, $X \in \C(\sphere,
\hat g_0 \cap \hat g^{\leq -i})$.
These functions are elements of the kernel $\Ker
\left( \{\, ,\, \}_R \right)$, and thus are constants of motion.
Fixing these constants of motion reproduces a Lax operator
with the form $\L =\partial_x + \Lambda +q(x), q(x) \in \C(\sphere,
\hat g_{\geq 0} \cap \hat g^{<i})$ and a gauge equivalence under
$\C(\sphere, P_-)$, \ie the generalised hierarchy of [\Tim].
The remaining constants of motion are those that are constructed from
the equivalence classes of $\hat u$. Choosing these constants of motion
selects a leaf of the foliation induced by the second Poisson bracket.

\chapter{ANOTHER VERY INTERESTING EXAMPLE}

In this section we complete
the analysis of the $Sl(3)$ theory associated with the
element $\Lambda= \Lambda_\alpha$.
We prove that under gauge reduction of the $G_\s$-orbit we can
reproduce the KdV hierarchy as defined in [\Tim].

In section 8, we constructed the $G_\s$-orbit, equation \Rorbits.
The fact that we have altered $I_+$ means that the gauge group is
different from that of section 8. In particular the symmetry group
$\tilde N_-$ is reduced from that in section 8 to the Lie group
generated by $\C(\sphere, \e)$. The momentum map takes the form
$$
J_\s(\q (Y))=
B^\prime-B^2 -2D,
$$
in terms of the parametrisation \Rorbits\ of $\C\left( \sphere,
\hat g_0 \mod \Ker(I_-) \right)$.
Thus we obtain the level set
constraint, momentum$=-2\mu$
$$
D= {1 \over 2} \left(B^\prime -B^2 \right)+\mu,
$$
with the level set
$$
\q=\Lambda_\alpha+\pmatrix{-B&0&0\cr -C&0&0 \cr
\left( B^\prime -B^2 \right) +\mu& A& B \cr}
\eqn\wphase
$$
Anticipating that the little group is again trivial, this form
for $\q$ will be a gauge slice for the theory. It is similar to the
Mason-Sparling gauge slice in [\MS].

The little group of this level set is calculated by solving for the
map $S$ of
definition 5, $S : \tilde N_- \rightarrow \C(\sphere,g \mod \Ker(I_-))$. The
calculation
$$
\pmatrix{1&0&0 \cr 0&1&0 \cr -\alpha &0&1 }
\left( \partial_x + I_+ \right)
\pmatrix{1&0&0 \cr 0&1&0 \cr \alpha &0&1 }=
\pmatrix{\alpha& 0& 1\cr 0&0&0 \cr \alpha^\prime -\alpha^2 & 0&-\alpha \cr}
$$
implies that
$$
S(e^{\alpha \e})=\pmatrix{{1 \over 2} \left(\alpha^\prime -\alpha^2
\right) &0& -\alpha \cr 0&0&0\cr 0&0&
-{1 \over 2} \left(\alpha^\prime -\alpha^2 \right)\cr}
$$
Thus the cocycle $\sigma(g)$ is given by
$\sigma(e^{\alpha \e})=2 \alpha^\prime$, which implies that
the little group is the constant subgroup of $\tilde N_-$.
Under $\Ad$-action the little group gauges away the
constant component of $B$, implying that this component is
non-dynamical.

In order to prove that the phase space, as parametrised in
\wphase, reproduces the KdV hierarchy defined in [\Tim], it is
necessary to prove that \wphase\ is a gauge slice for the theory.
The Lax operator of the hierarchy has an homogeneous grade zero
component with the general form, (before gauge fixing), [\Tim]
$$
\L_0=\partial_x+
\pmatrix{U^+ &J^+& 1 \cr G^+ &-U^+-U^-& J^-\cr T & G^- & U^- \cr}
\eqn\zero
$$
with a gauge equivalence under
$\C(\sphere, N_-)$. On the homogeneous grade zero component,
gauge equivalence reads
$$
\eqalign{
n_-^{-1} &\L_0 n_- =\cr \cr
&\pmatrix{U^+ +\a J^+ +\c &J^++ \b& 1 \cr
\hat G^+ & -U^+-U^- -\a J^+ +\b J^--\a\b& J^--\a \cr
\hat T
& \hat G^-
& U^--\c -\b J^- +\a\b \cr}
\cr}
\eqn\gaugeaction
$$
where $n_-$ has the form given in \thrgauge, and
$$\eqalign{
\hat T=&
T+\c^\prime-\b\a^\prime+\a G^--\b G^+ +\c (U^--U^+)+\a\b(2U^++U^-) + \cr
&\ \ \ \ \ \ \ \ \ \ \ \ \ \ \ \ \ \ \ \ \ \ \ \ \ \ \ \ \ \ \ \ \ \
\a^2 \b J^+-\c(\a J^++\b J^-)-\c^2 +\a\b\c \cr
\hat G^+=& G^+ +\a^\prime-\a \left(2U^++U^- \right)+\c J^- -\a\c-\a^2 J^+ \cr
\hat G^-=& G^- +\b^\prime +\b \left(U^++2U^-\right)-\c J^+
-\b\c-\b^2 J^-+\a\b J^++\a\b^2.
\cr}
\eqn\gaugevar
$$
Thus, through the gauge degrees of freedom we can remove the variables
$J^\pm$ and the component proportional to $H_1+H_2$,
\ie let $\a=J^-, \b=-J^+,
\c={1 \over 2} \left(U^- -U^+ -J^+ J^- \right)$. This gives a gauge slice
$$
\q=z \e+ \pmatrix{{1 \over 2} U & 0 & 1 \cr G^+ & -U &0 \cr T& G^- & {1
\over 2}U \cr}
\eqn\wtrad
$$
similar to the type used in the other examples of [\Tim].
We must prove that \wphase\ is also a good gauge slice. This follows
by performing a gauge transformation on \wphase\ such that it attains
the form \wtrad. Performing a gauge transformation from \wphase\ to \wtrad,
we find that
$T=2 B^\prime+\mu$. However, no components in the Cartan subalgebra
are reproduced, \ie we appear to have a
discrepancy in the treatment of $H_1-H_2$.
However, $H_1 -H_2 \in {\cal H}[w]$ and hence the field $U$
is non-dynamical, a field that can be
reproduced in \wphase\ by it's inclusion
in $\Lambda$. Thus, provided
$\int T(x) dx$ is non-dynamical, the phase space of the KdV hierarchy
is reproduced as the Hamiltonian reduction of a $G_\s$-orbit.
Calculating the first Hamiltonian of the theory through the
Drinfel'd-Sokolov proceedure, [\DS], we find that $H_{\Lambda_\alpha}
=\int T(x) dx$,
confirming that this fourier component of $T$ is a constant under all
the flows.

Consider the $G_R$-orbit of $\Lambda_\alpha$
$$
\q=z \e + g^{-1} \left( \partial_x+ \q_0 \right) g, \ g \in G.
$$
In order that this is dynamically equivalent
\Rorbits, it is necessary to
perform a gauge reduction as before, since this orbit contains
dynamical degrees of freedom that are not present in \Rorbits.
The momentum map is projection onto $n_+$ as before, \quotmap. However
the appropriate value of the momentum is $P_0^{\geq0}
\left( \Lambda_\alpha \right)$, which has a little group generated by
$\e$. Fixing this gauge freedom,
the Lax operator takes the form \wtrad, with the dynamical
variables $G^\pm, T$ parametrised by  $g \in G$
satisfying the level set constraints. Thus the phase space is foliated
by Hamiltonian reductions of $G_R$-orbits, reproducing the hierarchy
as a dynamical system.

\chapter{A FRACTIONAL KdV EXAMPLE: Sl(3)}

In this section we consider the case of a `fractional'
hierarchy with $\Lambda$ of ${\bf s}[w]$-grade 2. This example has
previously been studied in [\bakas, \burr, \Tim].
The fractional KdV hierarchy is defined by the Lax operator
$\L=\partial_x + \Lambda +q(x)$, where
$$
\Lambda=\pmatrix{0&0&1 \cr z &0&0 \cr 0&z &0 \cr}
$$
and $q(x) \in \hat g_{\geq0} \cap \hat g^{\leq 1}$, [\Tim]. The gauge group
is generated by $C^\infty(\sphere,\hat g_0 \cap \hat g^{<0})$.
In this section,
we prove that the two orbit constructions in section 4 reproduce
the $Sl(3)$ fractional KdV hierarchy as a dynamical system.

{}From our previous examples, we know that the
$G_\s$-orbit is more restictive as a model for the phase space, and
gives a gauge fixing proceedure.
We observe from [\Tim] that the components of the phase space with
homogeneous degree zero are identical to the example considered in the
previous section, \ie the Lax operator has an homogeneous degree zero
form identical to \zero\ before gauge fixing.
Thus, we use a realisation for $g \mod
\Ker(I_-)$ that produces a parametrisation of the $G_\s$-orbit
similar to \zero,
$$
Y=\pmatrix{{1 \over 3} \left(2G^+ +G^- \right) & -U^+ & -J \cr
{1 \over 2} T & -{1 \over 3} \left(G^+ -G^- \right) & U^- \cr
0& -{1 \over 2} T & -{1 \over 3} \left(G^+ +2 G^- \right) \cr}
$$
The orbit takes the form, equation \Iorbit
$$
\q= \Lambda + \pmatrix{U^+ & J& 0 \cr G^+ & -\left(U^+ +U^-\right)
& -J \cr T & G^- & U^-\cr},
\eqn\fractionorbit
$$
where we have not included a (non-dynamical) component proportional to
$z \e$ for simplicity. To reproduce the most general form of the
fractional KdV hierarchy, \ie reproducing all constants of motion, we
should include this component.
We observe that this orbit gives a Lax operator that is very similar to
the Lax operator constructed in [\Tim], {\it before} introducing gauge
equivalence. They
differ
principally in the symmetry of the components $J^\pm$, and
the fact that no component proportional to $z \e$ is created through
the $G_\s$-action. Introduction of
the gauge group in [\Tim] reduces the number of dynamical fields  to four,
a suitable gauge slice taking the form,
$$
\q=\pmatrix{0&0&1 \cr z&0&0 \cr 0&z&0 \cr}+
\pmatrix{{1 \over 2}U & 0&0\cr G^+& -U & 0 \cr T& G^- &{1 \over 2}U}
+ \phi \pmatrix{0&1&0 \cr 0&0&1 \cr z &0&0 \cr}
\eqn\fractionslice
$$
The field $\phi$ is not dynamical, and is in fact set to zero in
[\bakas].
In order to obtain equivalence between the integrable system
constructed
from the coadjoint orbit \fractionorbit\ and
the fractional KdV hierarchy defined by the Lax operator
\fractionslice, we must
reduce the degrees of freedom by 2.
This is achieved
through the Hamiltonian reduction of the orbit by the
gauge group $\tilde N_- \subset \C (\sphere, \hat g_0 \cap \hat
g^{<0})$, the subgroup that preserves the orbit.
This requires that $[X, I_-]=0, \forall X \in \tilde N_-$
such that the homogeneous degree 1 components of the Lax operator are
preserved. Thus $\tilde N_-$ consists of
matrices of the form \smallergauge.
{}From section 9, the momentum map
of the gauge group $ \tilde N_-$ is
$$
J_\s(\q(Y))=\left[\partial_x +I_+ , Y\right]+{1 \over 2}
\left[ \left[I_-, Y \right],Y \right]\  \mod \Ann(\tilde n_-),
$$
where $Y$ parametrises the orbit \fractionorbit.
This takes the form
$$\eqalign{
&J_\s(\q (Y))= \cr \cr
&\ \ \ \
-\pmatrix{*& U^{+\, \prime} +{1 \over 2}T +U^{+\, 2}+{1 \over 2} U^+
U^- &
J^\prime +J\left(U^+ - U^- \right) \cr
& +{1 \over 2}J\left(G^+-G^-\right)& +\left(G^+ + G^- \right) \cr
*&*& -U^{-\, \prime} +{1 \over 2}T +U^{-\, 2}+{1 \over 2} U^+
U^-\cr
&&+{1 \over 2}J\left(G^+-G^-\right)
\cr *&*&* \cr }
\cr}
$$
which gives the level set constraints
$$\eqalign{
J^\prime & +J\left(U^+ - U^- \right) +\left(G^+ + G^- \right)=0, \cr
T+ \left(U^+ -U^- \right)^\prime  & +U^{+\, 2}+U^{-\, 2}+ U^+
U^- +J\left(G^+-G^-\right)=0,\cr}
\eqn\twoconstraints
$$
where we have chosen the momentum map to have the value zero for the
convenience of discussion.
We anticipate that, as in previous examples, the little group is
effectively trivial, so that the Hamiltonian reduction by $\tilde N_-$
reduces the degrees of freedom by 2, through the 2 constraints in
\twoconstraints.
This is the correct number for
the theory to be equivalent to the fractional KdV hierarchy.
The
level set takes the form
$$\eqalign{
&\q= \Lambda+ \cr \cr
&\ \ \pmatrix{U^+ & J & 0 \cr
G- {1 \over 2} \left(J^\prime + J\left(U^+ -U^- \right) \right) &
-U^+ -U^- & -J \cr
\left(U^--U^+ \right)^\prime -U^{+\, 2} -U^{-\, 2} -U^+ U^-
-2JG &
-G- {1 \over 2} \left(J^\prime + J\left(U^+ -U^- \right) \right) &U^- \cr}
\cr}
\eqn\fractionlevel
$$
where we have defined $G= {1 \over 2} \left(G^+-G^- \right)$.
This level set possesses 4 dynamical degrees of freedom:
$U^+, U^-, G \equiv \left(G^+
-G^-\right)$ and $J$.
We propose
that this parametrisation of
the level set corresponds to a gauge slice of the fractional KdV
hierarchy of [\Tim].
To verify this, we must prove that the little group $G_\s(n_\s)$ is
`trivial', and that under a
full $C^\infty(\sphere, N_-)$ gauge transformation, the two slices
\fractionslice\ and \fractionlevel\ are equivalent.
The gauge equivalence relation for the homogeneous
degree zero component
is identical to that of
the previous example, and takes the form \gaugeaction.
The homogeneous grade 1 components are
invariant because $[X, I_-]=0, \forall X \in \tilde N_-$. We choose
$$
\a=\b=-J,\ \c={1 \over 2} \left(U^- -U^+ \right) +{1\over 2} J^2,
$$
in a gauge group action on \fractionlevel,
which produces a gauge slice
similar to \fractionslice, \ie
removing the ${\bf s}[w]$-grade
1 components, and the Cartan subalgebra component proportional
to $H_1 +H_2$.
The dynamical variables now take the form, \gaugevar\
$$\eqalign{
U=& \left(U^+ + U^- \right) -J^2,\cr
\hat G^+= & -J^\prime +G+ {3 \over 2}J \left(U^+ + U^- \right) - J^3,\cr
\hat G^- =& -J^\prime -G- {3 \over 2}J \left(U^+ + U^- \right) + J^3,\cr
\hat T=& -{3 \over 2} \left(U^+ -U^- \right)^\prime -{3 \over 4} U^2.\cr}
\eqn\relation
$$
Thus we have reproduced the gauge slice \fractionslice\ with the
constraints $\int G^+ +G^- dx=0, \int T +{3 \over 4}U^2 dx=0$.
By including a component $\phi \e$ into $I_-$, and
altering the value of the momentum map, we are able to reproduce the
gauge slice \fractionslice\ in all generality, with the priviso
that $\int G^+ +G^- dx, \int T +{3 \over 4}U^2 dx$ are non dynamical.
A calculation of the Hamiltonians of the theory through the
Drinfel'd-Sokolov proceedure, [\Tim, \DS] proves that this is in fact
the case, because $H_{\Lambda^1}=\int G^+ +G^-dx, H_{\Lambda^2}
=\int T+ {3 \over 4} U^2 dx$ where $\Lambda^1\equiv \Lambda_{{\rm
co}}, \Lambda^2 \equiv \Lambda_{{\rm co}}^2$ are the
elements of the Heisenberg subalgebra ${\cal H}[w]$ of ${\bf
s}[w]$-grades 1 and 2 respectively.

The final calculation that must be performed to complete the
proof of phase space
equivalence between the Hamiltonian reduction of the $G_\s$-orbit,
\fractionlevel,
and the fractional KdV hierachy is the calculation of the little
group, in particular the cocyle $\sigma$ of theorem 3.
This requires that we solve for the map
$S$, definition 5, given by
$$
g^{-1} \left( \partial_x + I_+ \right) g= I_+ + \left[I_-, S_g
\right],\  {\rm for\ } g= \pmatrix{1 &0&0 \cr \a & 1&0 \cr \c&\a&1 \cr}
\in \tilde N_-.
$$
Using the parametrisation given in equation \fractionorbit\ in terms
of the variable $Y$,
we deduce that
$$
S_g=\pmatrix{\a^\prime-\a\c+{1 \over 3}\a^3 & -\c &-\a \cr
{1 \over 2} \left(\c^\prime -\a\a^\prime -\c^2 +\c\a^2 \right) & {1 \over
3}\a^3 & -\c+\a^2 \cr
0 &-{1 \over 2} \left(\c^\prime -\a\a^\prime -\c^2 +\c\a^2 \right) &
-\a^\prime+\a\c-{2 \over 3} \a^3 \cr}
$$
This implies that the cocycle $\sigma(g^{-1})= \left[ \partial_x +I_+ ,
S_g \right]+ {1 \over 2} \left[ \left[ I_-, S_g\right] ,S_g \right]$ is
$$
\sigma(g^{-1})= \pmatrix{*& {1 \over 2} \left(-3\c^\prime +\a\a^\prime+\a^4
\right) -\c\a^2& -3\a^\prime \cr
*&*& {1 \over 2} \left( -3 \c^\prime +5\a\a^\prime -\a^4\right) +\c\a^2 \cr
*&*&*\cr}
$$
Since the gauge group $\tilde N_-$ is abelian, the little group
$G(n_\s)$ of all points $n_\s \in \tilde n_-$ is given by
$G(n_\s)=\{g \in \tilde N_-\, |\, \sigma(g)=0\}$, \ie
an element of $G(n_\s)$ satisfies the constraints
$\c^\prime=\a^\prime=0$. Thus the little group is the constant subgroup
of $\tilde N_-$, as in the previous examples. Under $\Ad$-action with
this little group, we obtain the equivalence relations
$$\eqalign{
U^+ \sim U^++\a J+\c,\  U^- \sim U^-+\a J+\a^2-\c, \cr
J\sim J+\a,\ G \sim G-{1 \over
2} \left(3\a \left(U^+ +U^-+\a J \right)+\a^2 \right), \cr}
\eqn\lileq
$$
on the fields $U^+, U^-, G, J$ of the level set \fractionlevel. Thus
the constant components of $J$ and $U^+-U^-$ can be gauged away.
It can be proved that there is no contradiction between this gauge
transformation and the parametrisation \relation, because the fields
$U,\hat G^\pm, \hat T$ of \relation\ are independent
of the equivalence relation  \lileq.

The $G_R$-orbit takes the form $ g^{-1} \L_0 g + \L_1$, and hence
contains dynamical degrees of freedom in excess of those contained in
the fractional KdV theory of [\bakas,\Tim].
Thus we introduce the gauge group and perform an Hamiltonian
reduction. However, as discussed in section 10,
$\Ad$-action by $\C(\sphere, N_-)$ is not a
symmetry of the orbit; $\Ad \left(\C(\sphere, N_-) \right)$ does not
stabilise the homogeneous grade 1 component, \ie $zI_-$.
The symmetry group is thus generated by $\e, e_{-1}+e_{-2}$. The
momentum map $J_R$ takes the value $I_+ \mod \Ann(h_R)$, with the
little group equal to the full symmetry group since the momentum map
is equivariant and the symmetry group abelian.
Through action by the little group, we obtain the gauge fixed form
$$
\q= \pmatrix{0&0&1 \cr z&0&0 \cr 0&z&0 \cr} +
\pmatrix{{1 \over 2}U & 0& 0 \cr G^+ & -U
& 0 \cr T & G^- & {1 \over 2}U\cr} + \pmatrix{0&\phi &0 \cr 0&0&  \phi
\cr z \hat \phi &0&0 \cr},
\eqn\ott
$$
where $\phi, \hat \phi$ are constants of motion, and $U= U^++U^-$.
There are 4 dynamical degrees of freedom
$U, G^\pm, T$, which are parametrised by
$g \in G$ satisfying the two level set constraints.
This gauge fixed form is identical to that of the fractional KdV
hierarchy of [\Tim], except for the treatment of the
constants of motion $\phi$ and $\hat \phi$.
The resolution of this can be accomplished in many ways. For example,
we can choose the momentum map to have the value $\phi =\hat \phi$, or
we can introduce the remaining part of the gauge group $\C(\sphere,
N_-)$ not included in $\tilde N_-$. This involves extending the
reduction to a Poisson manifold, and
removes us from the framework
of Hamiltonian reduction. It is well known that the $\Ad$-action by
$\C(\sphere, N_-)$ defines a Poisson symmetry of  $(\C(\sphere, \hat g_{\geq0}
\cap \hat g^{<i}), \{\, , \, \}_R)$, and hence there is a Poisson
structure induced on the gauge invariant functions, \ie the
space of gauge
equivalence classes of $\C(\sphere, \hat g_{\geq0} \cap \hat g^{<i})$
is a
Poisson manifold. We now prove that Lax operators of the form
\fractionslice\  are a
Poisson submanifold of this space, \ie the constraint $P^2 (\q)=I_+$
is consistent with the Poisson structure. This follows since $\hat
J_{\e}$ is both gauge invariant, and an element of the kernel $\Ker(\{\,
,\,\}_R)$. Hence $\hat J_{\e}$ is a constant of motion, the momentum
$\hat J_{\e}=1$ defines the submanifold consisting of
Lax operators of the form \fractionslice.
The further restriction to symplectic leaves can be accomplished by a
similar analysis, \ie the gauge group $\C(\sphere, N_-)$ preserves the
Poisson manifold $\{\q=g^{-1}\L_0 g + +zI_-+z \phi \e \, |\, \forall
g \in G, \phi \in
\C(\sphere) \}$, thus the equivalence classes under $\Ad (\C(\sphere,
N_-))$ also form a Poisson manifold, with a kernel generated by
the image of $\phi$ and $\hat J_{\e}$.

\chapter{Conclusions}

In this paper we have proved that there exists
a coadjoint orbit structure for the phase spaces
of the KdV hierachies associated
to $Sl(2)$ and $Sl(3)$, $i<N$. Two group actions are constructed,
$G_R$, $G_\s$, such that the Poisson structures of [\burr] are reproduced
under the Kirillov construction, [\kirillov].
We reconstruct the phase space $\cal M$ of the generalised hierarchies
of [\Tim], $i<N$, as an Hamiltonian reduction of a coadjoint orbit
of the $G_\s$-action. The gauge group $H_\s$ is the maximal subgroup
of $\C(\sphere, N_-)$ that
preserves the orbit.
The Hamiltonian reduction by this symmetry is
non-trivial, the momentum map being equivariant with respect
to an extended $\Ad$ action. However, we observe that the little group
is effectively trivial, hence the calculation of the level set gives a
gauge fixing of the theory. This gauge slice is a generalisation of
the gauge slice
employed in [\MS]. If this extends  to all the hierarchies,
this process would
solve the problem of how to choose a gauge slice for the generalised
hierarchies of [\Tim].
The orbits of the $G_R$-action
produce a partition of the potentials by `type', this being finer than
the partition into mKdV, pmKdV and `true' KdV type potentials. The
fact that the phase space $\cal M$ cannot be reproduced from a single
orbit is a result of the fact that the Poisson structure $\{\, , \,
\}_2 \equiv \{\, ,\, \}_R$ possesses a non-trivial kernel on $\cal M$.

This paper has only considered the problem of reproducing the phase
space of the KdV hierarchies from a coadjoint orbit method. In
particular we choose the gauge groups such that the theories in [\Tim]
are reproduced, and use certain observations about constants of
motion to verify equivalence. Ultimately, the coadjoint orbit
construction should be an independent and self contained construction.
We propose that given $I_-$,
the symmetry group $H_\s$, (which also defines $I_+$),
is specified by the requirement that
the reduced space is  foliated by Hamiltonian
reductions of $G_R$-orbits.
The gauge group $H_R$ is identical for each
$G_R$-orbit of interest.
It then remains to prove that the Poisson
brackets are coordinated, and further that the theory is
bi-Hamiltonian. It is expected that the regularity of
the element $\Lambda$  will be reproduced as a method of classification.
{}From our examples we observe that the symmetry group $H_\s$ restricts
the possible form of $I_+$ given $I_-$.
We hope to consider these questions in a later publication.

This construction of the generalised KdV hierarchies of [\Tim] from a
Coadjoint Orbit framework suggests many avenues for further
generalisations. We observe that the orbits used in this construction
are a special case, suggesting that there may exist a further
generalisation involving characters, as in the full AKS theory,
[\babelon]. We have also only considered untwisted Kac
Moody algebras, and restricted the potentials to be periodic.

Finally we remark that the $G_{R[w]}$-action discussed in sections
6 and 10 is very reminiscient of
the generation of solutions to the KdV hierarchies
through dressing transformations,
[\DJ,\Jim,\SW]. The construction involves a vertex operator
representation of a Kac Moody algebra, and generates tau-functions
from the vacuum.
The use of the vertex representation is suspected to be
analogous to our restriction
to level sets of the $G_R$-orbits.
The connection between coadjoint orbits and unitary representations,
[\kirillov], supports the  suggestion of a relationship.
In this vein, we further observe that
the Poisson
structure analysis of the dressing transformation in
[\shansky] is very suggestive of the
$G_{R[w]}$-action discussed in sections 6 and 10, and it's associated
Poisson mapping structure.

\ack

I would like to thank Tim Hollowood for many valuable discussions on the
contents of this paper.
The research in this paper was supported by grant
\#DE-FG02-90ER40542.

\refout

\bye